\documentclass{article}
\usepackage[letterpaper,top=2cm,bottom=2cm,left=3cm,right=3cm,marginparwidth=1.75cm]{geometry}
\usepackage{graphicx} 
\usepackage{natbib}
\usepackage{bm}
\usepackage{amssymb}
\usepackage[colorlinks=true, allcolors=blue]{hyperref}
\usepackage{amsthm}
\usepackage{amsmath}
\usepackage{authblk}
\usepackage{caption}
\usepackage{subcaption}
\usepackage{booktabs}

\title{Model-Agnostic Influential Outlier Detection for Mixed Effects and Multi-Level Models}
\author{Colin C. Jones\\ Institute for Data Science, Carleton University \and David A. Campbell\thanks{davecampbell@cunet.carleton.ca}\\ School of Mathematics and Statistics\\ and School of Computer Science, Carleton University \and Yan Liu\\ Department of Psychology, Carleton University}
\date{\today}

\begin{document}

\maketitle
\begin{abstract}
Influential Outlier Detection is developed for mixed-effects models on clustered data.  The Influential Outlier Metric is defined as a combination of SHapley Additive exPlanantion (SHAP) values and model residuals, both of which undergo a change of measure transformation. Building on previous work showcasing the suitability of using Normalizing flows to map arbitrary distributions to a flexible base distribution for statistical inference, the Normalizing Flows are constructed to allows contextual information and also provide a goodness of fit diagnostic for model evaluation.  The use of SHAP values in the construction moves away from model specific tools and instead provides point-wise model agnostic influential outlier. The advantages and limitations of this approach are examined in several models including the linear model, the random forest, and gradient-boosted trees. 
\end{abstract}

\section{Introduction}
Outlier detection remains a fundamental concern in statistical modeling and predictive analytics. Outliers are observations that depart markedly from the dominant structure of the data. While not all outliers are problematic, those that meaningfully alter model behavior are of particular interest and are commonly referred to as influential observations. Failure to properly identify and assess such points can lead to degraded predictive performance and misleading inferential conclusions \citep{Aggarwal2017}.

To formalize this setting, consider a predictive framework in which the response vector $\mathbf{y} \in \mathbb{R}^n$ is related to a set of $p$ covariates $\mathbf{X} \in \mathbb{R}^{n \times p}$ through an unknown function $f(\cdot)$ with parameter vector $\bm{\theta}$ and error term $\bm{\varepsilon} \in \mathbb{R}^n$, such that
\begin{equation}\label{definition}
\mathbf{y} = f(\mathbf{X}; \bm{\theta}) + \bm{\varepsilon}.
\end{equation}
Within this general formulation, influential observations are those whose presence induces a non-negligible change in the fitted function $f(\cdot)$, thereby affecting both predictions and parameter estimates.

When $f(\cdot)$ is specified as a linear function, the concept of influence is well understood and can be quantified through Cook’s distance \citep{Cook1977}. This measure leverages the geometry of the design matrix via the hat matrix, whose diagonal elements $h_{ii}$ capture the leverage of individual observations. Cook’s distance for observation $i$ can be written as
\begin{equation}\label{cd-2}
C_i = \left[\frac{y_i - \mathbf{x}_i' \hat{\bm{\beta}}}{\hat{\sigma}\sqrt{1 - h_{ii}}}\right]^2 \frac{h_{ii}}{p(1 - h_{ii})},
\end{equation}
where $\hat{\bm{\theta}} = \{\hat{\sigma}, \hat{\bm{\beta}}\}$ are the Ordinary Least Squares (OLS) estimators of the error standard deviation and regression coefficients and $\mathbf{a}'$ represents the transpose of vector $\mathbf{a}$. This formulation highlights that influence arises from the interaction between two distinct components: the square of the residual, which reflects the discrepancy between observed and fitted values, and the leverage, which quantifies how unusual an observation is in the covariate space.

The Influential Outlier Metric (IOM)  provides model-agnostic influential outlier detection in fixed effect models, including machine learning (ML) algorithms such as neural networks and bagged/boosted trees. The influence of an observation in an ML model uses a combination of SHapley Additive exPlanantion (SHAP) values \citep{Lundberg2017}  and the residual. Both are transformed using normalizing flows (NF) \citep{Papamakarios2021}, changing their respective  measures to Gaussian distributions.  The transformed SHAP values and residuals define the IOM as a product of two independent $\chi^2$ random variables. Formally, define the model-agnostic IOM as the product

\begin{equation}
\mathrm{IOM}_i = \left(\langle\bm{\Phi}\rangle_Z - \bm{\mu}_{\bm{\Phi}}\right)_i'\bm{\Sigma}_{\bm{\Phi}}^{-1}\left(\langle\bm{\Phi}\rangle_Z - \bm{\mu}_{\bm{\Phi}}\right)_i\left(\frac{\langle\hat{\varepsilon}\rangle_{Z i} - \mu_{\hat{\varepsilon}}}{\sigma_{\hat{\varepsilon}}}\right)^2,
\end{equation}where $\langle\cdot\rangle_Z$ indicates the inverse transformation of a NF to change the measure of the SHAP values and residuals to a Gaussian random variable ($\langle\bm{\Phi}\rangle_Z\sim \mathcal{N}(\bm{\mu}_{\bm{\Phi}}, \bm{\Sigma}_{\bm{\Phi}})$ and $\langle\hat{\varepsilon}\rangle_{Zi}\sim \mathcal{N}(\mu_{\hat{\varepsilon}}, \sigma_{\hat{\varepsilon}}^2)$). Since this is the product of independent $\chi^2_p$ and $\chi^2_1$ random variables, the density \citep{joarder2011statistical} of $\mathrm{IOM}$ is 

\begin{equation}
f_{\mathrm{IOM}}(x)
=
\frac{x^{(p+1)/4 - 1}}
     {2^{(p+1)/2 - 1}\sqrt{\pi}\Gamma(p/2)}
K_{(p-1)/2}(\sqrt{x}),
\qquad x > 0 ,
\end{equation}where $\Gamma(\cdot)$ is the gamma function and $K_{\nu}(\cdot), \nu \geq 0$ is the modified Bessel function of the second kind.

An observation is flagged as an influential outlier at significance level $\alpha$ when
\begin{equation}
\int_{\xi_{p,\alpha}}^\infty f_{\mathrm{IOM}}(x) \, \mathrm{d}x = 1 - \alpha ,
\end{equation}
where $\xi_{p,\alpha}$ is the upper-tail cutoff.

In this paper, the established framework is extended to substantially more flexible modeling environments. The key contribution of this paper is the development of the Mixed-Effects Influential Outlier Metric (MEIOM), which extends influential outlier detection to mixed-effects machine learning models. We integrate this perspective within mixed-effects machine learning models \citep{hajjem2014mixed, olaniran2025mixed}, which jointly capture complex nonlinear relationships and hierarchical dependence structures. 

Further, new functionality comes from improvements to the normalizing flow (NF) methodology used for change-of-measure inference. First, we generalize the unimodal Gaussian base distribution used in IOM to settings in which the latent distribution is better characterized by a mixture of Gaussian distributions \citep{Papamakarios2021}. This extension allows the NF to accommodate multimodal and more complex data topologies that may otherwise require substantial transport to map to a single Gaussian base distribution. This enables a wider range of SHAP value and residual distributions to be represented while also providing a new diagnostic tool for identifying latent multimodal structure. Second, we extend the NF to allow observed contextual information to condition the transformation when such information is available. To support this increased flexibility, we develop an information criterion for selecting the appropriate number of mixture components by balancing the complexity of the latent base distribution against the amount of transport required by the NF.

The remainder of this paper is structured as follows. Section~\ref{methods} introduces mixed-effects models and classical influence diagnostics as well as mixed-effect SHAP values, residuals. Next, Section~\ref{MEIOM} develops the MEIOM. Section~\ref{NF-extensions} extends the normalizing-flow framework to Gaussian-mixture base distributions and conditional transformations, while Section~\ref{IOM_mix} extends the IOM to the mixture-Gaussian setting. Section~\ref{NFIC} introduces the Normalizing Flow Information Criterion (NFIC) for base-distribution selection, and Section~\ref{verification} demonstrates its use for flexible normalizing-flow base distributions. Section~\ref{example} evaluates the proposed methodology in a simulated hierarchical setting, and Section~\ref{application} applies the MEIOM to a real-world PISA dataset. Finally, Section~\ref{conclusion} concludes.

\section{Mixed-Effects Models}\label{methods}
In order to extend influence diagnostics to clustered data, consider a mixed-effects model (MEM) with a random intercept, defined by a response vector $y_{ij} \in \mathbb{R}$, individual-level (level-1) fixed effect predictors $\mathbf{x}_{ij}\in \mathbb{R}^p$, cluster-level (level-2) fixed effect predictors $\mathbf{w}_j \in \mathbb{R}^r$, a random intercept vector $b_j \in \mathbb{R}$, an error term $\varepsilon_{ij} \in \mathbb{R}$, with $i=1,\dots, n_j$, $j=1,\dots, q$ and $n=n_1+\cdots+n_q$. Along with a model $f$ with parameters $\boldsymbol{\theta}$, the MEM is organized by the relation,

\begin{equation}
    y_{ij} = f(\mathbf{x}_{ij}, \mathbf{w}_{j}; \bm{\theta}) + b_j + \varepsilon_{ij}.
\end{equation}For the linear MEM, $f(\mathbf{x}_{ij}, \mathbf{w}_j; \bm{\theta}) = \mathbf{a}_{ij}\bm{\beta}$ and,

\begin{equation}\label{lme}
    y_{ij} = \mathbf{a}_{ij}\bm{\beta} + b_j + \varepsilon_{ij}
\end{equation}where $\mathbf{a}_{ij}=[\mathbf{x}_{ij}\,\mathbf{w}_j]$ and $\bm{\theta}=\bm{\beta}$. One may further suppose for the linear mixed effect model that $\varepsilon_{ij} \sim \mathcal{N}(0, \sigma^2)$, $b_j \sim \mathcal{N}(0, \sigma^2_b)$ and $\mathrm{Cov}(\varepsilon_{ij}, b_j)=0$.

\subsection{Mixed-Effects Cook's Distance}
In terms of influence, a full leave-one-out (or leave-one-group-out) approach to Cook's distance is applied for the hierarchical levels (level 1 $(i)$ and level 2 $(j)$) due to the added complexity of the random effects. Following \citet{christensen1992case}, there exists two sets of diagnostics: fixed-effects Cook's distance and random-effects Cook's distance. First for the fixed effects for level 1,

\begin{equation}
    C_i(\hat{\bm{\beta}}) = \frac{1}{p}(\hat{\bm{\beta}}-\hat{\bm{\beta}}_{(i)})'\widehat{\mathrm{Var}(\hat{\bm{\beta}})}^{-1}(\hat{\bm{\beta}}-\hat{\bm{\beta}}_{(i)})
\end{equation}where $\hat{\bm{\beta}}_{(i)}$ indicates the estimate of the fixed-effect parameters with the $i^{th}$ individual removed. And for level 2,

\begin{equation}
    C_j(\hat{\bm{\beta}}) = \frac{1}{p}(\hat{\bm{\beta}}-\hat{\bm{\beta}}_{(j)})'\widehat{\mathrm{Var}(\hat{\bm{\beta}})}^{-1}(\hat{\bm{\beta}}-\hat{\bm{\beta}}_{(j)})
\end{equation}where $\hat{\bm{\beta}}_{(j)}$ indicates the estimate of the fixed-effect parameters with the $j^{th}$ group removed.

For the random effects let $\boldsymbol{\zeta} = [\sigma^2, \sigma_b^2]'$, for level 1,

\begin{equation}
    R_i(\hat{\boldsymbol{\zeta}}) = (\hat{\boldsymbol{\zeta}}-\hat{\boldsymbol{\zeta}}_{(i)})'\widehat{\mathrm{Var}(\hat{\boldsymbol{\zeta}})}^{-1}(\hat{\boldsymbol{\zeta}}-\hat{\boldsymbol{\zeta}}_{(i)})
\end{equation}where $\hat{\boldsymbol{\zeta}}_{(i)}$ indicates the estimate of the random-effect parameters with the $i^{th}$ individual removed. And for level 2,

\begin{equation}
    R_j(\hat{\boldsymbol{\zeta}}) = (\hat{\boldsymbol{\zeta}}-\hat{\boldsymbol{\zeta}}_{(j)})'\widehat{\mathrm{Var}(\hat{\boldsymbol{\zeta}})}^{-1}(\hat{\boldsymbol{\zeta}}-\hat{\boldsymbol{\zeta}}_{(j)})
\end{equation}where $\hat{\boldsymbol{\zeta}}_{(j)}$ indicates the estimate of the random-effect parameters with the $j^{th}$ group removed. Note there is no normalizing constant for random effect case-deletion diagnostics (see \citep{christensen1992case}).

\subsection{Mixed-effect SHAP values and residuals}\label{ME_shap_resid}
SHAP (SHapley Additive exPlanations) \citep{Lundberg2017} values quantify the contribution of each input feature to a model's prediction, providing a consistent and interpretable decomposition of the prediction into feature-level effects. SHAP values can be used to form an additive, linear, and local explanation model $g$ to approximate the complex regression model $f$. Function $g(\bm{\varphi})$ is defined over surrogate variables, $\bm{\varphi}$, used to locally interpret the impact of variables $\mathbf{x}_{ij}$ and $\mathbf{w}_j$. To account for population-level and group-specific variation in MEMs, SHAP values and residuals can be calculated for fixed-effect components (at all levels), random-effect components, and residual components.

\subsubsection{SHAP values}
For mixed-effects SHAP values, the local explanation model $g$ can be formulated as,

\begin{equation}\label{shap}
f(\mathbf{x}_{ij}, \mathbf{w}_{j}) \approx g(\varphi_{ij}) = \varphi_{0i} + \sum_{k=1}^p\varphi_{ki} + \sum_{j=p+1}^r\varphi_{j}.
\end{equation}Next, denote $\bm{\Phi}^{(i)}$ as the collection of SHAP values for the level-1 fixed effects and $\bm{\Phi}^{(j)}$ as SHAP values for the level-2 fixed effects. The construction of the MEIOM combines these mixed effect SHAP values, along with the associated MEM residuals introduced in the next section.

\subsubsection{Residuals}
\citet{hilden1995multilevel} report four types of residuals for MEMs: level-1 (conditional) residuals $\hat{\varepsilon}_{ij}$, level-1 marginal residuals $\hat{e}_{ij}$, the estimated random intercept as level-2 residuals, $\hat{b}_j$ from equation \eqref{lme}, and level-2 marginal residuals $\bar{e}_j$. 

The level-1 conditional residuals,
\begin{equation}\label{cond_resid}
\hat{\varepsilon}_{ij} =  y_{ij} -  f(\mathbf{x}_{ij}, \mathbf{w}_{j}; \hat{\bm{\theta}}) - \hat{b}_j,
\end{equation} are the variation that is leftover after the individual, group fixed effects and the random group-specific deviation from those fixed effects are accounted for. For example, this determines how well the model explains this student's score after adjusting for the effect of their school. The level-1 marginal residuals,

\begin{equation}
e_{ij} = y_{ij} - f(\mathbf{x}_{ij}, \mathbf{w}_{j}; \hat{\bm{\theta}}),
\end{equation} measure how well the fixed effects explain the outcome on average, ignoring individual/group variation. For example, this determines how well age and gender predict test scores, ignoring which school the student is in.

The level-2 residuals, $\hat{b}_j$ are  estimated from the regression equation capturing unobserved cluster-level factors that systematically increase or decrease the outcome across all observations within a group. The level-2 marginal residuals are

\begin{equation}
\begin{split}
\bar{e}_j &= 1/n_j\sum_{i=1}^{n_j}y_{ij} - 1/n_j\sum_{i=1}^{n_j} f(\mathbf{x}_{ij}, \mathbf{w}_{j}; \hat{\bm{\theta}}) \\
&= \bar{y}_j - \bar{f}(\mathbf{x}_{ij}, \mathbf{w}_{j}; \hat{\bm{\theta}}),
\end{split}\label{eq:lvl2resid}
\end{equation}
representing how much the average outcome for a group differs from what the fixed-effects portion of the model predicts for that group.

\section{Mixed effect influential outlier metric}\label{MEIOM}
In this section, the the mixed-effect influential outlier metric (MEIOM) is developed for both individuals (level 1) and groups (level 2). 

\subsection{MEIOM: Level 1 ($i$)}
The conditional residuals  capture the direct effect of each individual on the model. Let the NF transform the SHAP values \eqref{shap} and conditional residuals \eqref{cond_resid}  into Gaussian base distributions ($\langle \cdot \rangle_Z$, described in Section \ref{NF-extensions}).  The level-1 MEIOM is formulated by the product of the transformed level-1 SHAP values and  transformed level-1 conditional residuals,

\begin{equation}
\mathrm{MEIOM}_i^{Z} = (\langle \bm{\Phi}^{(i)} \rangle_{Z} - \bm{\mu}_{\bm{\Phi}^{(i)}})' \bm{\Sigma}_{\bm{\Phi}^{(i)}}^{-1} (\langle \bm{\Phi}^{(i)} \rangle_{Z} - \bm{\mu}_{\bm{\Phi}^{(i)}})_i \left(\frac{\langle \hat{\varepsilon}\rangle_{{Z}i}-\mu_{\hat{\varepsilon}}}{\sigma_{\hat{\varepsilon}}}\right)^2.
\end{equation}

\subsection{MEIOM: Level 2 ($j$)}
At the second level, the MEIOM can be defined using the transformed group-level SHAP values \eqref{shap} and the transformed estimated random intercept level-2 residual, $\langle\hat{b}_j\rangle_Z \sim \mathcal{N}(\mu_{\hat{\mathbf{b}}},\sigma^2_{\hat{\mathbf{b}}})$,

\begin{equation}
\mathrm{MEIOM}^{Z}_j = (\langle \bm{\Phi}^{(j)} \rangle_{Z} - \bm{\mu}_{\bm{\Phi}^{(j)}})' \bm{\Sigma}_{\bm{\Phi}^{(j)}}^{-1} (\langle \bm{\Phi}^{(j)} \rangle_{Z} - \bm{\mu}_{\bm{\Phi}^{(j)}})_j \left(\frac{\langle \hat{b}\rangle_{{Z}j}-\mu_{\hat{\mathbf{b}}}}{\sigma_{\hat{\mathbf{b}}}}\right)^2.
\end{equation}

Following this definition, a NF mapping ensures that both level-1 and level-2 MEIOMs are calibrated to a known reference distribution, facilitating principled inference and thresholding for influential observations. However, with more complex data topologies a mapping to a distribution other then a unimodal Gaussian can be advantageous. Further, conditional NF may provide more context to the flow to achieve more flexible mapping.

\section{Extending the normalizing flows for inference}\label{NF-extensions}
Normalizing flows use a sequence of invertible and differentiable transformations to map observations from a complex distribution to a simpler base distribution with a known density. Consider a random vector $\mathbf{V}$ from a complex or unknown distribution $p_{\mathbf{V}}(\mathbf{v}; \bm{\theta})$ and a simple base distribution $p_\mathbf{Z}(\mathbf{z}; \bm{\theta}_\mathbf{Z})$ A NF, $T$, maps $\mathbf{V} = T(\mathbf{Z})$. Both $\mathbf{V}$ and $\mathbf{Z}$ must have the same dimension, $p$. Under these conditions, the density of $\mathbf{V}$ is defined by the change-of-variable formula,

\begin{equation}
    p_{\mathbf{V}}(\mathbf{v}; \bm{\theta}) = p_\mathbf{Z}(\mathbf{z}; \bm{\theta}_\mathbf{Z})|\det J_T(\mathbf{z})|^{-1},
\end{equation}
with the $p\times p$ Jacobian,

\begin{equation}
    J_T(\mathbf{z})=
    \begin{bmatrix}
        \frac{\partial T_1}{\partial z_1} & \dots & \frac{\partial T_p}{\partial z_1}\\
        \vdots & \ddots & \vdots \\
        \frac{\partial T_p}{\partial z_1} & \dots & \frac{\partial T_p}{\partial z_p}
    \end{bmatrix}.
\end{equation}
In practice, $T$ is constructed with a neural network and may be composed of $K$ transformations:
\begin{equation}
     T = T_K \circ T_{K-1} \circ \dots \circ T_1, 
\end{equation}and the parameters of $T$ are estimated by minimizing the penalized loss function $\mathcal{L}$,

\begin{equation}
   \hat{T} = \underset{T}{\arg \min} \ \mathcal{L} = \underset{T}{\arg\min} \ \mathcal{L}_{KL} + \lambda \mathcal{L}_\mathrm{pen},
\end{equation}where,

\begin{equation}\label{KL}
    \begin{split}
        \mathcal{L}_{KL}(\bm{\theta}, \bm{\theta}_\mathbf{Z}) & = D_{KL}\left[ p^*_{\mathbf{V}}(\mathbf{v}) \| p_{\mathbf{V}}(\mathbf{v}; \bm{\theta})\right]\\
         & \propto -\mathbb{E}_{p^*_{\mathbf{V}}(\mathbf{v})}\left[\log p_{\mathbf{V}}(\mathbf{v}; \bm{\theta}))\right] \\
         & \propto -\mathbb{E}_{p^*_{\mathbf{V}}(\mathbf{v})}\left[\log p_\mathbf{Z}(T^{-1}(\mathbf{v}; \bm{\theta}); \bm{\theta}_\mathbf{Z})) + \log |\det J_{T^{-1}}(\mathbf{v}; \bm{\theta})|\right]\\
    \end{split}
\end{equation}and

\begin{equation}\label{penalty}
    \mathcal{L}_\mathrm{pen} = \mathbb{E}[\log^2 |\det J_{T^{-1}}(\mathbf{v})|].
\end{equation}

For the transformation to be useful for inference, it must provide a robust and stable change of measure. \citet{jonescampbell} examine the stability of NF transformations with respect to random initialization, sampling variability, and network architecture, providing empirical support for their use as a reliable inferential transformation when the mapped distribution satisfies the required goodness-of-fit conditions. Following \citet{ryzhikov2021nfad}, \citet{jonescampbell} use a penalty (parameterized by $\lambda>0$) on the distortion on the volume, calculated by the square of the determinant of the Jacobian matrix, done by the NF to restrict the movement away from the original data. Building on the stability analysis done by \citet{jonescampbell}, the framework is extended in two ways. First, a more flexible Gaussian-mixture base distribution is developed, together with an appropriate testing procedure to assess whether the transformed data are adequately calibrated to the chosen mixture. Second, a conditional NF is introduced, allowing observed contextual information to influence the transformation.

\subsection{Mixture of Gaussians base distribution}
Let $\mathbf{V} \sim p_{\mathbf{V}}(v)$ denote either the SHAP vectors or the residual vectors. The goal is to apply a transformation so that the inversely mapped variables $\langle \mathbf{V} \rangle_{\mathbf{Z}}$ follow a Gaussian mixture of the form

\begin{equation}
\langle \mathbf{V} \rangle_{\mathbf{Z}} \sim \sum_{k=1}^K w_k \, \mathcal{N}_p(\boldsymbol{\mu}_k, \boldsymbol{\Sigma}_k), 
\qquad \sum_{k=1}^K w_k = 1,
\end{equation}providing a more versatile reference distribution. To verify that the transformation is correctly specified, a goodness-of-fit test is developed for the base mixture model.

\subsubsection{Mixture-Gaussian Calibration by Mode Assignment and Whitening}\label{sec:test}
It is necessary to verify that the normalizing flow mapping adequately matches the base distribution based on goodness-of-fit-tests. The component means $\bm{\mu}_k$ and covariances $\bm{\Sigma}_k$ are chosen so that the modes of the mixture are well separated, ensuring that for each observation, $\exists k^*\in\{1,\dots,K\}$ such that $w_{k^*}\approx1$ and $w_{\ell\not=k^*}\approx 0$,  reducing the mixture to $\mathbf{v}_i \overset{.}{\sim}\mathcal{N}_p(\boldsymbol{\mu}_{k^*}, \boldsymbol{\Sigma}_{k^*})$. 

Conditional on the assigned component, each observation is whitened using the corresponding component mean and covariance matrix. For diagonal covariance matrices, this reduces to the coordinate-wise transformation
\begin{equation}
r_{ij}=\frac{V_{ij}-\mu_{k^*_i,j}}{\sigma_{k^*_i,j}},\qquad j=1,\dots,p.
\end{equation}If the transformed data are correctly calibrated to the target mixture distribution and the component assignment is reliable, then within each assigned mode the whitened residuals should satisfy

\begin{equation}
\mathbf{r}_i \mid k_i=k_i^*
\overset{.}{\sim}
\mathcal{N}_p(\mathbf{0},\mathbf{I}_p).
\end{equation}

This condition is evaluated using mode-level normality tests. In the univariate case, $p=1$, the Jarque-Bera test is used within the assigned mode. When $p\geq 2$, the Henze-Zirkler test for multivariate normality. This gives one mode-level $p$-value, denoted $p_\ell$, for each mixture component with sufficient assigned observations.

Because the mode-level tests assess local calibration of each component, they are reported individually. To summarize the aggregate evidence across components, Stouffer's combined statistic is also calculated,
\begin{equation}
Z_{\mathrm{St}}
=
\frac{1}{\sqrt{k^\ast}}
\sum_{\ell=1}^{k^\ast}
\Phi_F^{-1}(1-p_\ell),
\end{equation}
where $k^\ast$ is the number of components for which a mode-level normality test is computed and $\Phi_F$ is standard normal cumulative distribution function. The corresponding combined $p$-value is
\begin{equation}
p_{\mathrm{St}}
=
1-\Phi_F(Z_{\mathrm{St}}).
\end{equation}
The $p$-value $p_{\mathrm{St}}$ acts as an aggregate diagnostic of mode-level normality rather than as a replacement for the individual component tests. This is important because a combined $p$-value can mask a localized failure in one component. Therefore, the individual mode-level $p$-values and the Stouffer combined $p$-value are interpreted jointly.

The combined whitened residuals 
\begin{equation}
Q_i
=
\mathbf{r}_i^\prime \mathbf{r}_i
=
\sum_{j=1}^{p} r_{ij}^2,
\end{equation}
under correct calibration and reliable component assignment has distribution
\begin{equation}
Q_i \overset{.}{\sim}\chi^2_p.
\end{equation}
This pooled chi-square approximation is tested using goodness-of-fit tests such as the Kolmogorov-Smirnov test and the Cramér-von Mises test. This diagnostic is complementary to the mode-level normality tests: the mode-level tests assess whether each assigned component has the correct Gaussian shape, while the pooled chi-square test assesses whether the overall squared-distance scale is calibrated.

The penalty parameter $\lambda$ is selected using the goodness-of-fit tests as diagnostics rather than by relying on a single combined test. For each candidate regularization strength $\lambda$, compute the mode-level normality $p$-values, the Stouffer combined $p$-value, the assignment-certainty diagnostics, and the pooled chi-square goodness-of-fit $p$-value. A candidate transformation is considered acceptable only if no component shows severe lack of normality, the aggregate mode-level normality diagnostic is adequate, and the pooled chi-square statistic is not significantly non chi-square.

\subsection{Conditional normalizing flows}\label{sec:conditional_flow}
In settings where the distribution of the SHAP values or residuals is expected to vary systematically across an observed contextual variable, the normalizing-flow transformation is extended to a conditional NF. Let $\mathbf{C}_i$ denote the observed context for observation $i$. Rather than estimating a single marginal transformation $T^{-1}(\mathbf{V}_i)$, the conditional flow estimates
\begin{equation}
    \mathbf{Z}_i = T^{-1}(\mathbf{V}_i\mid \mathbf{C}_i),
\end{equation}
where $\mathbf{V}_i$ denotes either the SHAP vector or the residual and $\mathbf{C}_i$ is treated as fixed conditioning information rather than as a variable to be transformed. For each fixed value of $\mathbf{C}_i$, the map remains invertible with respect to $\mathbf{V}_i$.

The conditional transformation is implemented using a Neural Spline Flow (NSF) \citep{Durkan2019}. In an autoregressive rational-quadratic spline layer, the conditioning network that produces the spline widths, heights, and derivatives receives both the transformed variables available under the autoregressive mask and the context $\mathbf{C}_i$. Consequently, the spline transformation can vary with the observed context while sharing a common set of neural-network parameters across all observations. The context therefore modifies the transformation but is not itself included in the flow Jacobian. The change-of-variables density is
\begin{equation}
    p_{\mathbf{V}}(\mathbf{v}_i\mid \mathbf{C}_i)
    = p_{\mathbf{Z}}\!\left(T^{-1}(\mathbf{v}_i\mid \mathbf{C}_i)\right)
    \left|\det\left(\frac{\partial T^{-1}(\mathbf{v}_i\mid \mathbf{C}_i)}{\partial \mathbf{v}_i}\right)\right|.
\end{equation}

Importantly, the latent base distribution is kept common across contexts. Thus, conditioning affects the transport map but not the Gaussian or Gaussian-mixture reference distribution. This allows observations from different contexts to be mapped to the same calibrated latent distribution while defining outlyingness relative to the distribution expected within the observed context. When no contextual variable is supplied, the method reduces to the unconditional normalizing-flow formulation described above. Without loss of generality,  assume that the NF $\langle \cdot \rangle_{\mathbf{Z}}$ may be given a contextual variable if available $\langle \cdot \rangle_{\mathbf{Z}|\mathbf{C}}$.

\section{Mixed Effects Influential Outlier Metric with a Mixture Gaussian Base Distribution}\label{IOM_mix}

The IOM is extended to the case where the base distribution is a mixture of
Gaussians with fixed weights, means, and covariance matrices. Following the
single Gaussian base case, the Mahalanobis distance of the mixture of Gaussians
is of interest,
\begin{equation}
Q = \sum_{k=1}^K w_k \,
(\langle\mathbf{V}\rangle_\mathbf{Z} - \bm{\mu}_k)'
\bm{\Sigma}_k^{-1}
(\langle\mathbf{V}\rangle_\mathbf{Z} - \bm{\mu}_k),
\end{equation}
where $\langle\mathbf{V}\rangle_\mathbf{Z} \sim \sum_{k=1}^K w_k \,
\mathcal{N}_p(\bm{\mu}_k, \bm{\Sigma}_k)$. In general, this weighted sum of
squared Mahalanobis distances follows a weighted mixture of non-central
chi-square distributions, which can be complex to work with analytically.
However, given well-separated modes verified in Section~\ref{sec:test}, $Q$
can be approximated by a single $\chi^2_p$.

In what follows, each observation is assigned to the mixture component $k^*$
with the highest posterior probability, and all distances are computed with
respect to the parameters of that component.

\subsection{Level-1 MEIOM}\label{IOM_mix_l1}

Beginning with the transformation of the level-1 SHAP values \eqref{shap},
compute the squared Mahalanobis distance
\begin{equation}
D_{i}^2 =
\left(\langle \bm{\Phi}^{(i)} \rangle_{\mathbf{Z}i} - \bm{\mu}_{\bm{\Phi}^{(i)}_{k^*}}\right)'
\bm{\Sigma}_{\bm{\Phi}^{(i)}_{k^*}}^{-1}
\left(\langle \bm{\Phi}^{(i)} \rangle_{\mathbf{Z}i} - \bm{\mu}_{\bm{\Phi}^{(i)}_{k^*}}\right),
\end{equation}
where $D_{i}^2 \overset{.}{\sim} \chi^2_p$.

Similarly, apply the mixture of Gaussian base distribution to the level-1
conditional residuals \eqref{cond_resid} and take the transformed residual 
\begin{equation}
E_{i} =
\frac{\langle \hat{\varepsilon} \rangle_{\mathbf{Z}i} -
\mu_{\hat{\varepsilon}_{k^*}}}{\sigma_{\hat{\varepsilon}_{k^*}}}.
\end{equation}
If the modes are well separated, the approximation
$E_{i}^2 \overset{.}{\sim} \chi^2_1$ is valid; otherwise, $E_{i}^2$ is a
mixture of non-central $\chi^2$ distributions.

The level-1 influential outlier metric combines the Mahalanobis distance with
the normalized conditional residual:
\begin{equation}
\mathrm{MEIOM}^{\mathbf{Z}}_i = D_{i}^2 \cdot E_{i}^2.
\end{equation}

\subsection{Level-2 MEIOM}\label{IOM_mix_l2}

Analogously, for the level-2 SHAP values \eqref{shap} compute
\begin{equation}
D_{j}^2 =
\left(\langle \bm{\Phi}^{(j)} \rangle_{\mathbf{Z}i} - \bm{\mu}_{\bm{\Phi}^{(j)}_{k^*}}\right)'
\bm{\Sigma}_{\bm{\Phi}^{(j)}_{k^*}}^{-1}
\left(\langle \bm{\Phi}^{(j)} \rangle_{\mathbf{Z}i} - \bm{\mu}_{\bm{\Phi}^{(j)}_{k^*}}\right),
\end{equation}
where $D_{j}^2 \overset{.}{\sim} \chi^2_r$, and take the transformed level-2
estimated random intercept residual
\begin{equation}
E_{j} =
\frac{\langle \hat{b} \rangle_{\mathbf{Z}j} - \mu_{\hat{\mathbf{b}}_{k^*}}}{\sigma_{\hat{\mathbf{b}}_{k^*}}},
\end{equation}
where, again, $E_{j}^2 \overset{.}{\sim} \chi^2_1$ provided the modes are
well separated.

The level-2 influential outlier metric is then
\begin{equation}
\mathrm{MEIOM}^{\mathbf{Z}}_j = D_{j}^2 \cdot E_{j}^2.
\end{equation}

At both levels, since $D_{i}^2 \overset{.}{\sim} \chi^2_p$, $D_{j}^2 \overset{.}{\sim} \chi^2_r$, and $E_{\ell}^2 \overset{.}{\sim} \chi^2_1$ for $\ell \in \{i, j\}$, critical values
can be obtained from the product of independent $\chi^2$ distributions, as in
the single Gaussian case.

\paragraph{Summary of Steps:}
\begin{enumerate}
\item Choose means and diagonal covariance matrices such that mixture modes are well
separated. With $\mathbf{V} \sim \mathcal{N}_p(\bm{\mu}, \bm{\Sigma})$, modes separated by 
\begin{equation}
    \Delta_{k,\ell} = \sqrt{\sum_{j=1}^{p} \left(\frac{\mu_{kj}-\mu_{\ell j}}{\sigma_j}\right)^{2}} \geq 5
\end{equation} heuristically work well.
\item Compute posterior probabilities $\hat{\gamma}_{\ell k}, \ell\in \{i,j\}$ for each individual or school.
\item Assign each observation to the component $k^*$ with the highest
$\hat{\gamma}_{\ell k},  \ell\in \{i,j\}$.
\item Compute $D_{\ell}^2, $ and $E_{\ell}^2$ for level $\ell \in \{i,j\}$ with
respect to the assigned $k^*$.
\item Construct the level-1 and level-2 metrics as
$\mathrm{MEIOM}^{\mathbf{Z}}_\ell = D_{\ell}^2 \cdot E_{\ell}^2$, $\ell\in \{i,j\}$.
\item Use critical values from the product of independent $\chi^2$
distributions to identify influential outliers at each level.
\end{enumerate}

\section{Latent Structure Identification via the Normalizing Flow Information Criterion}\label{NFIC}
The flexibility of the mixture base distribution creates a model-selection problem: additional components may improve fit but can also introduce unnecessary latent complexity. Therefore select among candidate base distributions using the NFIC, counting only the free parameters of the mixture base (component means, diagonal scales, and mixture weights), while excluding the flow parameters. The flow parameters are optimized separately and screened by the preceding goodness-of-fit tests. Beyond model selection, the NFIC can also serve as a diagnostic tool. In particular, a residual distribution that strongly favours a multimodal base may indicate latent heterogeneity, omitted structure, or model misspecification that is not captured by the fitted model.

Let $\{\mathbf{v}_i\}_{i=1}^{n_v}$ denote the validation sample of either SHAP values or residuals, and for candidate flow $m$ let $\mathbf{z}_{im}=\hat{T}_m^{-1}(\mathbf{v}_i)$ be the corresponding latent representations under the fitted inverse map $\hat{T}_m^{-1}$. We introduce the \emph{Normalizing Flow Information Criterion} (NFIC), a transport-based criterion derived from a generalized Bayesian (Gibbs) evidence argument. The derivation proceeds in three steps: a Gibbs-style generalized evidence built from the transport loss; a Laplace approximation; and extraction of the leading sample-size-dependent terms to obtain the BIC-style penalty.

\subsection{Gibbs Evidence Motivation}

To derive a principled penalty for model complexity, consider the empirical transport loss $M_m(\bm{\theta}_m)$ with parameter vector $\bm{\theta}_m$,
\begin{equation}
M_m(\bm{\theta}_m)=\frac{1}{n_v}\sum_{i=1}^{n_v}l_i(\bm{\theta}_m),
\end{equation}
where $l_i(\bm{\theta}_m)$ is the transport cost for validation observation $i$. A Gibbs likelihood based on this loss is $L_G(\bm{\theta}_m) = \exp\{-n_v M_m(\bm{\theta}_m)/(2\tau^2)\}$, with scale parameter $\tau^2>0$, and the corresponding generalized evidence is
\begin{equation}
G_m = \int \exp\!\left\{-\frac{n_v}{2\tau^2}M_m(\bm{\theta}_m)\right\} p(\bm{\theta}_m\mid m)\, \mathrm{d}\bm{\theta}_m,\label{eq:Gm}
\end{equation}
where $p(\bm{\theta}_m\mid m)$ denotes the prior density on $\bm{\theta}_m$ under model $m$. By analogy with classical Bayesian model selection, $-2\log G_m$ will play the role of an information criterion; the NFIC arises by approximating this quantity in the large-$n_v$ regime.

\paragraph{Choice of the temperature parameter.}
The parameter $\tau^2$ acts as a temperature in the Gibbs likelihood, controlling concentration around $\hat{\bm{\theta}}_m$ without changing the transport-loss minimizer. To simplify (\ref{eq:Gm}), set $\tau^2=1$ because a common positive scaling does not affect model rankings. This choice yields the standard BIC-like penalty $\log n_v/n_v$. $\tau^2$ only temporarily written to show its role in the Laplace approximation.

\subsection{Laplace Approximation}

To extract a tractable expression for $G_m$, apply a Laplace approximation around the transport-loss minimizer $\hat{\bm{\theta}}_m = \arg\min_{\bm{\theta}_m} M_m(\bm{\theta}_m)$, and let $\mathbf{A}_m = \nabla^2 M_m(\hat{\bm{\theta}}_m)$ denote the Hessian at the minimum. A second-order Taylor expansion gives
\begin{equation}
M_m(\bm{\theta}_m) \approx M_m(\hat{\bm{\theta}}_m) + \frac{1}{2}(\bm{\theta}_m-\hat{\bm{\theta}}_m)' \mathbf{A}_m (\bm{\theta}_m-\hat{\bm{\theta}}_m),
\end{equation}
which isolates the optimized transport loss $M_m(\hat{\bm{\theta}}_m)$: the term that will survive as the data-fit component of the NFIC. Substituting into the generalized evidence yields
\begin{equation}
G_m \approx \int \exp\!\left\{-\frac{n_v}{2\tau^2}\!\left[M_m(\hat{\bm{\theta}}_m) + \frac{1}{2}(\bm{\theta}_m-\hat{\bm{\theta}}_m)' \mathbf{A}_m (\bm{\theta}_m-\hat{\bm{\theta}}_m)\right]\right\} p(\bm{\theta}_m\mid m)\, \mathrm{d}\bm{\theta}_m .
\end{equation}

Assuming $p(\bm{\theta}_m\mid m)$ is smooth and positive in a neighborhood of $\hat{\bm{\theta}}_m$,  approximate $p(\bm{\theta}_m\mid m)\approx p(\hat{\bm{\theta}}_m\mid m)$ and pull the prior outside the integral, so that
\begin{equation}
G_m \approx \exp\!\left\{-\frac{n_v}{2\tau^2}M_m(\hat{\bm{\theta}}_m)\right\} p(\hat{\bm{\theta}}_m\mid m) \int \exp\!\left\{-\frac{n_v}{4\tau^2}(\bm{\theta}_m-\hat{\bm{\theta}}_m)' \mathbf{A}_m (\bm{\theta}_m-\hat{\bm{\theta}}_m)\right\}\mathrm{d}\bm{\theta}_m.
\end{equation}

The remaining integral is Gaussian in $\bm{\theta}_m$ and captures the local curvature of the transport loss around $\hat{\bm{\theta}}_m$. Here, the number of mixture parameters, $K_m$, enters because the integral is taken over a $K_m$-dimensional parameter space. For any positive-definite $K_m\times K_m$ matrix $\mathbf{B}_m$,
\begin{equation}
\int
\exp\!\left\{-\tfrac{1}{2}\mathbf{u}'\mathbf{B}_m \mathbf{u}\right\}\mathrm{d}\mathbf{u}
=
(2\pi)^{K_m/2}|\mathbf{B}_m|^{-1/2}.
\end{equation}

In the Laplace approximation, the precision matrix is
\begin{equation}
\mathbf{B}_m=\frac{n_v}{2\tau^2}\mathbf{A}_m,
\end{equation}
where $\mathbf{A}_m$ is the Hessian of the transport loss at $\hat{\bm{\theta}}_m$. Therefore,

\begin{align}
\int
\exp\!\left\{
-\frac{n_v}{4\tau^2}
(\bm{\theta}_m-\hat{\bm{\theta}}_m)'\mathbf{A}_m (\bm{\theta}_m-\hat{\bm{\theta}}_m)
\right\}
\mathrm{d}\bm{\theta}_m
&=
(2\pi)^{K_m/2}
\left|
\frac{n_v}{2\tau^2}\mathbf{A}_m
\right|^{-1/2}  \\
&=
(2\pi)^{K_m/2}
\left(\frac{n_v}{2\tau^2}\right)^{-K_m/2}
|\mathbf{A}_m|^{-1/2},
\end{align}which leaves us with,

\begin{equation}
G_m \approx \exp\!\left\{-\frac{n_v}{2\tau^2}M_m(\hat{\bm{\theta}}_m)\right\}p(\hat{\bm{\theta}}_m\mid m)
(2\pi)^{K_m/2}
\left(\frac{n_v}{2\tau^2}\right)^{-K_m/2}
|\mathbf{A}_m|^{-1/2}.
\end{equation}

\subsection{Asymptotic Form}

To isolate the NFIC penalty, extract the leading terms in $n_v$. Taking logarithms gives
\begin{equation}
\log G_m \approx -\frac{n_v}{2\tau^2}M_m(\hat{\bm{\theta}}_m) + \log p(\hat{\bm{\theta}}_m\mid m) + \frac{K_m}{2}\log(2\pi) - \frac{K_m}{2}\log\!\left(\frac{n_v}{2\tau^2}\right) - \frac{1}{2}\log |\mathbf{A}_m|.
\end{equation}
The first term scales linearly in $n_v$ and the fourth scales as $K_m\log n_v$; the remaining terms are bounded as $n_v\to\infty$. Retaining only sample-size-dependent terms yields
\begin{equation}
-2\log G_m \approx \frac{n_v}{\tau^2}M_m(\hat{\bm{\theta}}_m) + K_m\log n_v + O(1),
\end{equation}
which already exhibits the BIC-like trade-off between fit and complexity that underpins the NFIC. Dividing by $n_v$ produces the per-observation form
\begin{equation}
\frac{-2\tau^2}{n_v}\log G_m \approx M_m(\hat{\bm{\theta}}_m) + \tau^2\frac{K_m\log n_v}{n_v} + O\!\left(\frac{1}{n_v}\right).
\end{equation}
At this point the role of $\tau^2$ is fully transparent: it multiplies the penalty term but does not interact with the parameter dimension $K_m$, the sample size $n_v$, or the optimized transport loss $M_m(\hat{\bm{\theta}}_m)$. Setting $\tau^2 = 1$, as motivated above, yields,

\begin{equation}
-\frac{2}{n_v}\log G_m \approx M_m(\hat{\bm{\theta}}_m) + \frac{K_m\log n_v}{n_v} + O\!\left(\frac{1}{n_v}\right),
\end{equation}
in which the data-fit term and the penalty $K_m\log n_v / n_v$ stand on equal footing, exactly the structure carried forward into the NFIC.

\subsection{The NFIC Criterion}

Dropping the asymptotically negligible $O(1/n_v)$ remainder, the unit-temperature per-observation expression above defines the criterion
\begin{equation}
\mathrm{NFIC}_m = M_m + \rho K_m, \qquad \rho=\frac{\log n_v}{n_v}.
\end{equation}
Thus the NFIC inherits its goodness-of-fit term directly from the optimized transport loss in the Gibbs evidence, while its penalty $\rho K_m$ arises from the Laplace-approximation determinant factor evaluated at unit temperature.

It remains to specialize $K_m$ to the present model class. For a diagonal Gaussian mixture base with $P_m$ components in $p$ dimensions, the number of free base-distribution parameters is
\begin{equation}
K_m = \underbrace{P_m p}_{\text{means}} + \underbrace{P_m p}_{\text{scales}} + \underbrace{(P_m-1)}_{\text{free weights}} = 2P_m p + (P_m-1).
\end{equation}
Combining this parameter count with the empirical squared-error transport distance $M_m = n_v^{-1}\sum_{i=1}^{n_v}\|\mathbf{v}_i-\mathbf{z}_{im}\|_2^2$ gives the final operational form of the criterion,
\begin{equation}
\mathrm{NFIC}_m = \frac{1}{n_v}\sum_{i=1}^{n_v}\left\|\mathbf{v}_i-\mathbf{z}_{im}\right\|_2^2 + \frac{\log n_v}{n_v}\bigl\{2P_m p+(P_m-1)\bigr\},
\end{equation}
which is evaluated for each candidate model that survived the goodness-of-fit tests; the model attaining the smallest $\mathrm{NFIC}_m$ is selected.

\section{Demonstrations of Flexible Normalizing Flow Base Distributions}\label{verification}
\citet{jonescampbell} demonstrate that the NF transformation used in the original IOM is appropriately stable with respect to random initialization, sampling variability, and network architecture, supporting its use as a robust change-of-measure technique for inference. Building on these results, we examine whether additional flexibility in the latent base distribution can improve the transformation when the underlying data topology is more complex. In particular, we show that for certain multimodal structures, a bimodal Gaussian base distribution provides a better representation than a unimodal alternative, as determined by the NFIC.

Consider a random sample, $X_1,\dots, X_{1000}$, drawn from the following distribution,

\begin{equation}
X_i = Z_i + 0.500 \sin(1.40 Z_i) + 0.150 \sin(3Z_i)
\end{equation} where

\begin{equation}
    Z_i \sim 0.600\,\mathcal{N}(-3.20, 0.550^2) + 0.400\,\mathcal{N}(2.80, 1.25^2).
\end{equation}

\begin{figure}
    \centering
    \includegraphics[width=0.500\linewidth]{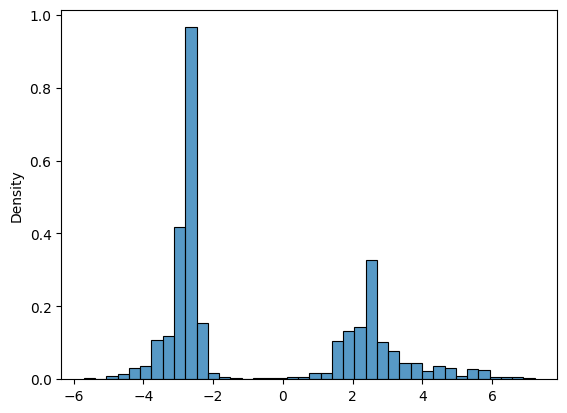}
    \caption{Distribution of $X$ with complex data topology}
    \label{fig:warped_bimodal}
\end{figure}

\begin{figure}
    \centering

    \begin{subfigure}[t]{0.310\textwidth}
        \centering
        \includegraphics[width=\textwidth]{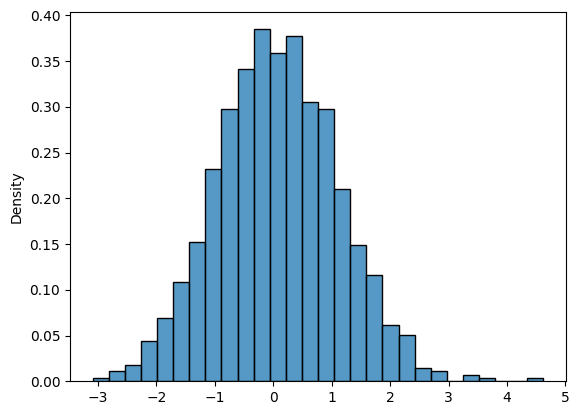}
        \caption{Unimodal base after inverse transformation}
        \label{fig:unimodal_base}
    \end{subfigure}%
    \hfill
    \begin{subfigure}[t]{0.310\textwidth}
        \centering
        \includegraphics[width=\textwidth]{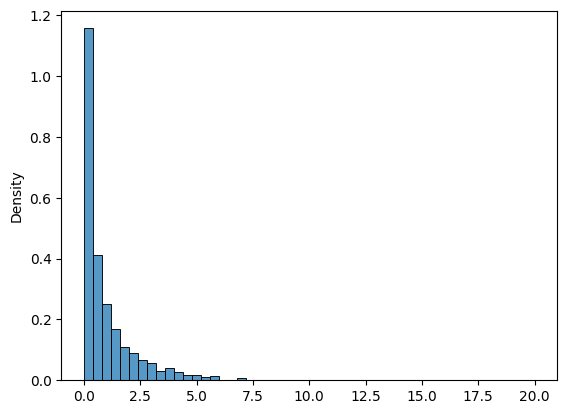}
        \caption{Transformed $\chi^2$ distribution}
        \label{fig:unimodal_chi2}
    \end{subfigure}%
    \hfill
    \begin{subfigure}[t]{0.310\textwidth}
        \centering
        \includegraphics[width=\textwidth]{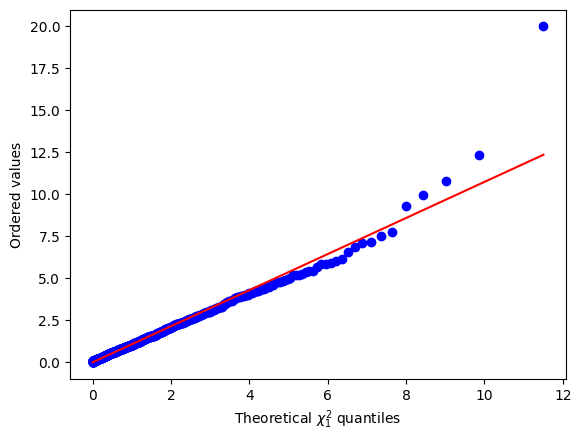}
        \caption{QQ plot of transformed $\chi^2$ distribution}
        \label{fig:unimodal_qq}
    \end{subfigure}

    \caption{Unimodal base distribution}
    \label{fig:unimodal}
\end{figure}

\begin{figure}
    \centering

    \begin{subfigure}[t]{0.310\textwidth}
        \centering
        \includegraphics[width=\textwidth]{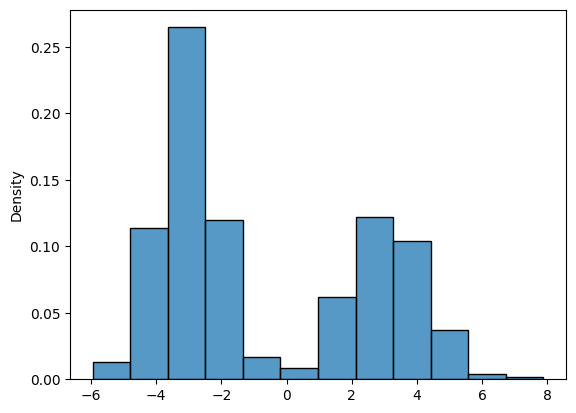}
        \caption{Bimodal base after inverse transformation}
        \label{fig:bimodal_base}
    \end{subfigure}%
    \hfill
    \begin{subfigure}[t]{0.310\textwidth}
        \centering
        \includegraphics[width=\textwidth]{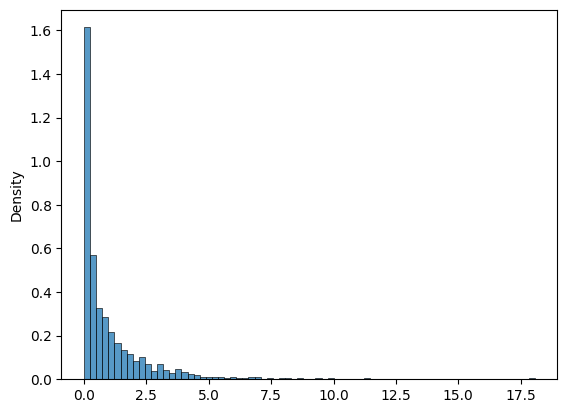}
        \caption{Transformed $\chi^2$ distribution}
        \label{fig:bimodal_chi2}
    \end{subfigure}%
    \hfill
    \begin{subfigure}[t]{0.310\textwidth}
        \centering
        \includegraphics[width=\textwidth]{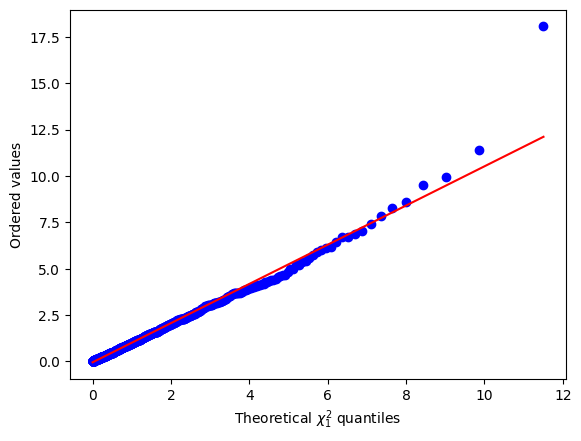}
        \caption{QQ plot of transformed $\chi^2$ distribution}
        \label{fig:bimodal_qq}
    \end{subfigure}

    \caption{Bimodal base distribution}
    \label{fig:bimodal}
\end{figure}

\begin{figure}
    \centering

    \begin{subfigure}[t]{0.310\textwidth}
        \centering
        \includegraphics[width=\textwidth]{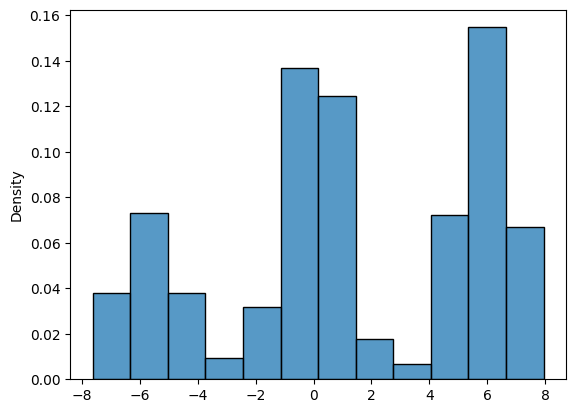}
        \caption{Trimodal base after inverse transformation}
        \label{fig:trimodal_base}
    \end{subfigure}%
    \hfill
    \begin{subfigure}[t]{0.310\textwidth}
        \centering
        \includegraphics[width=\textwidth]{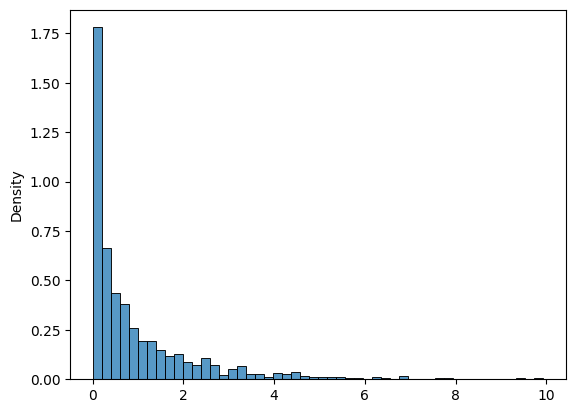}
        \caption{Transformed $\chi^2$ distribution}
        \label{fig:trimodal_chi2}
    \end{subfigure}%
    \hfill
    \begin{subfigure}[t]{0.310\textwidth}
        \centering
        \includegraphics[width=\textwidth]{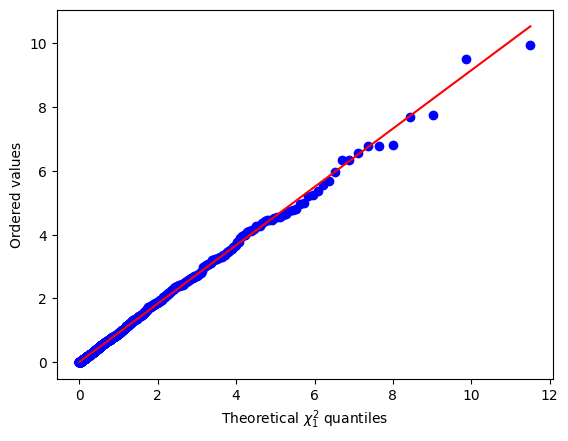}
        \caption{QQ plot of transformed $\chi^2$ distribution}
        \label{fig:trimodal_qq}
    \end{subfigure}

    \caption{Trimodal base distribution}
    \label{fig:trimodal}
\end{figure}

The NF was fitted using Neural Spline Flow (NSF) \citep{Durkan2019} with three different Gaussian mixtures: unimodal (Figure \ref{fig:unimodal}), bimodal (Figure \ref{fig:bimodal}) and trimodal (Figure \ref{fig:trimodal}). Across all mixtures the learning rate (0.000300), the number of epochs (1000), the validation split (80/20), the tail bound (8) and ability to train base parameters (True) are kept constant. The confidence level for the Jarque-Bera tests was set to be 0.01, the Stouffer test to be 0.05 and the final Cram\'er-von Mises (CvM) test to be 0.05. The data is centered to have a zero mean but not scaled to be of unit variance. This means that the tail bound parameter and number of bins parameter need to be adjusted to account for the non-unit variance and the wigglyness of the mixtures. They are chosen so that the data does not reject the distribution checks and are regularized by the $\lambda$ parameter. The parameter specification and test results are shown in Table \ref{tab:verifcation_params}. 

Table \ref{tab:verifcation_params},  Figures \ref{fig:unimodal} through \ref{fig:trimodal}, as well as the Stouffer $p$-value show that the modes are well separated and there is no reason to reject normality for each mode. The final CvM test and QQ plot do not reject the $\chi^2_1$ fit.  The NFIC strongly favours the bimodal specification: The bimodal model has the lowest NFIC value, $2.60$, compared with $4.78$ for the unimodal model and $8.73$ for the trimodal model. Since lower NFIC values indicate a more favourable balance of transport simplicity and base-distribution parsimony among likelihood-competitive models, the bimodal specification is selected by the proposed criterion.

\begin{table}
    \centering
    \begin{tabular}{lccc}
    \toprule
         &  \textbf{Unimodal} & \textbf{Bimodal} & \textbf{Trimodal} \\
        \hline
        $\mu_i$ & 0.0353 & $\{-2.99, 3.08\}$ & $\{-5.71, -0.0263, 5.92\}$ \\
        $\sigma_i$ & 1.03 & $\{0.938, 1.13\}$ & $\{1.11, 0.953, 0.933\}$ \\
        $w_k$ & 1 & $\{0.597, 0.403\}$ & $\{0.241, 0.385, 0.374\}$ \\
        Transformations & 10 & 6 & 8 \\
        Hidden units & 10 & 6 & 8 \\
        Hidden layers & 2 & 2 & 2 \\
        Number of bins & 20  & 14 & 32 \\
       $\lambda$ &  0.00670 & 0.0131 & 0.0256 \\
       $\hat{p}_{\text{JB}}$ & 0.0833 & $\{0.998, 0.0162\}$ & $\{0.0972, 0.963, 0.291\}$ \\
       $\hat{p}_{\text{St}}$ & 0.0833 & 0.691 & 0.486 \\
       $\hat{p}_{\text{CvM}}$ & 0.880 & 0.710 & 0.338 \\
       NFIC & 4.78 & 2.60 & 8.73 \\
       \bottomrule
        
    \end{tabular}
    \caption{Parameter values, $p$-values and NFIC for the three mixture distributions.}
    \label{tab:verifcation_params}
\end{table}

\section{Example}\label{example}
This example demonstrates the MEIOM in a controlled hierarchical setting where influential observations are introduced at both the student and school levels. A mixed-effects data-generating process is constructed with known student- and school-level outliers and a bimodal residual distribution, allowing us to evaluate both the hierarchical extension of the IOM and the more flexible NF base distributions developed above. First one can compare the identified influential observations with classical fixed- and random-effects Cook's distance and then apply the MEIOM using both a linear MEM and a Mixed Effects Random Forest (MERF).

\subsection{Data Generation}\label{data_generation}
A hierarchical dataset is generated with $n_{\text{sch}} = 100$ schools and 
$n_{\text{stu}} = 30$ students per school, for a total of 
$n = n_{\text{sch}} \times n_{\text{stu}}$ observations. 
Each student $i$ belongs to a school $j \in \{1,\dots,100\}$.

\paragraph{Predictors.}
Each student has two predictors. The first predictor $x_{ij}$ is drawn from a 
Gaussian distribution:
\begin{equation}
x_i \sim \mathcal{N}(0, 1);
\end{equation}
the school level predictor is constructed similarly:
\begin{equation}
w_j \sim \mathcal{N}(0, 1).
\end{equation}

\paragraph{Random intercepts.}
Schools are assigned random intercepts:
\begin{equation}
b_j \sim \mathcal{N}(0, 1).
\end{equation}

\paragraph{Bimodal residual distribution.}
Idiosyncratic errors $\varepsilon_{ij}$ are drawn from a two-component Gaussian mixture:
\begin{equation}
\varepsilon_{ij} \sim 
0.600\,\mathcal{N}(3, 1.20^2) \;+\; 
0.400\,\mathcal{N}(-3, 1.20^2).
\end{equation}

\paragraph{Outcome model.}
The outcome is generated using a linear mixed-effects structure:
\begin{equation}
y_{ij} = 2 + 1.50\,x_{ij} + 0.750\,w_j + b_j + \varepsilon_{ij}.
\end{equation}

\paragraph{School-level outliers.}
Designate $n_{\text{sch,out}} = 5$ schools as \emph{outlier schools}, 
selected randomly without replacement. For each outlier school $s \in \mathcal{S}_{\text{out}}$, the random intercept is augmented with with the addition of the following vector

\begin{equation}
b_j = [2,4,6,8,10]' \quad \text{for each } s  \in \mathcal{S}_{\text{out }} \text{respectively}.
\end{equation} Given the standard normal distribution of $b_j$ we expect that $b_j=2$ will not be identified by the MEIOM whereas $b_j=4$ may be identified and $b_j=6,8,10$ should be identified. The school-level variable $w$ is also set at the following

\begin{equation}
w_j = 4 \quad \text{for all } s  \in \mathcal{S}_{\text{out} }.
\end{equation}

\paragraph{Student-level outliers.}
Introduce $n_{\text{stu,out}} = 3$ student-level outliers, sampled only from 
non-outlier schools. For each selected student $r\in \mathcal{I}_{\text{out}}$, overwrite both predictor 
and outcome values:
\begin{equation}
x_{ij} = 4, \quad y_{ij} = 16, \quad \text{for all } r \in \mathcal{I}_{\text{out}}. 
\end{equation}

\begin{figure}
    \centering
    \includegraphics[width=0.500\linewidth]{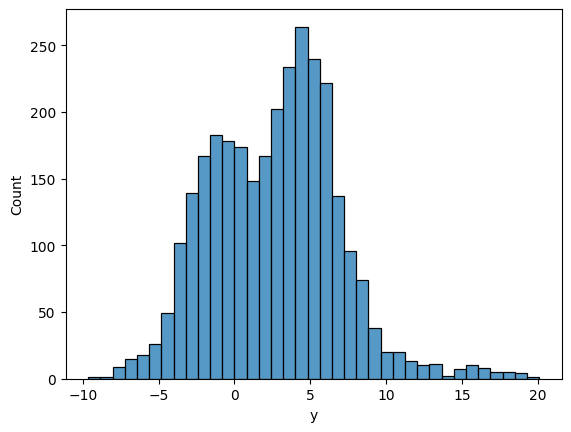}
    \caption{Distribution of the outcome variable with and outliers, and bimodal residuals.}
    \label{fig:y}
\end{figure}

\paragraph{Final output.}
This design produces a challenging DGP with cluster-level contamination, observation-level contamination, and a bimodal residual distribution. Figure~\ref{fig:y} plots the distribution of the outcome variable.

\subsection{Cook's distance}
As a comparison, the linear mixed effects model \eqref{lme} is fitted to the data generated in Section \ref{data_generation}. The labeled outlier students are $\mathcal{I}_{\text{out}}=\{1028, 1602, 451\}$ whereas the labeled outlier schools are $\mathcal{S}_{\text{out}}=\{90, 18, 60, 94, 67\}$. 

\begin{figure}
 \begin{subfigure}[b]{0.490\textwidth}
         \includegraphics[scale=0.350]{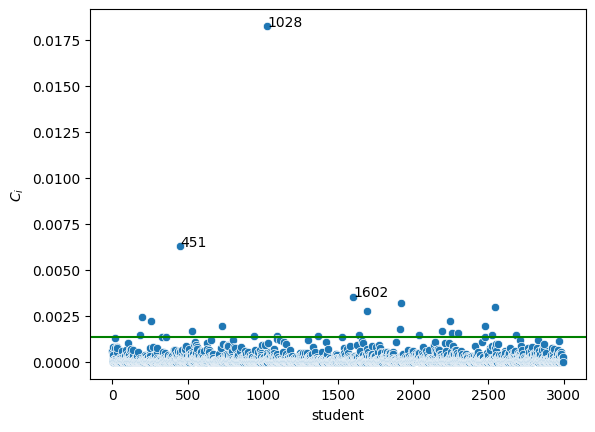}
    \caption{Influential students based on fixed-effects Cook's distance with green line as the rule-of-thumb cutoff.}
        \label{fig:c_i} 
     \end{subfigure}
 \begin{subfigure}[b]{0.490\textwidth}
    \includegraphics[scale=0.350]{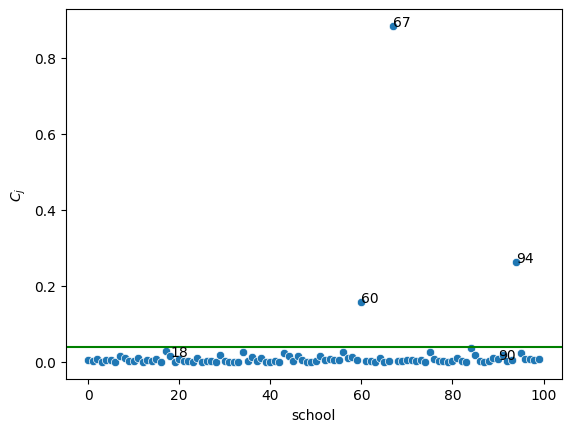}
    \caption{Influential schools based on fixed-effects Cook's distance with green line as the rule-of-thumb cutoff.}
    \label{fig:c_j} 
     \end{subfigure}
\caption{Cook's distance based on fixed effects}
\label{fig:C}
\end{figure}

\begin{figure}
 \begin{subfigure}[b]{0.490\textwidth}
         \includegraphics[scale=0.350]{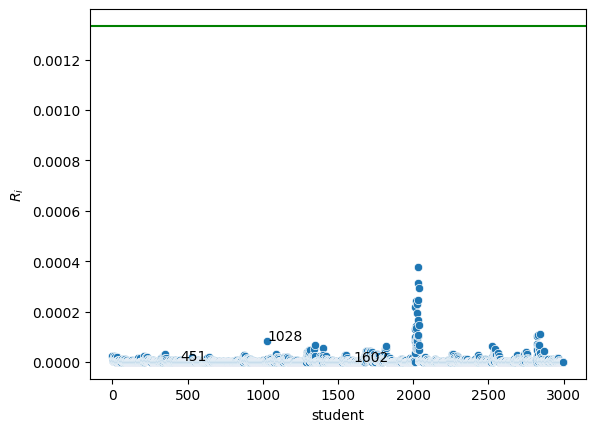}
    \caption{Influential students based on random-effects Cook's distance with green line as the rule-of-thumb cutoff.}
        \label{fig:r_i} 
     \end{subfigure}
 \begin{subfigure}[b]{0.490\textwidth}
    \includegraphics[scale=0.350]{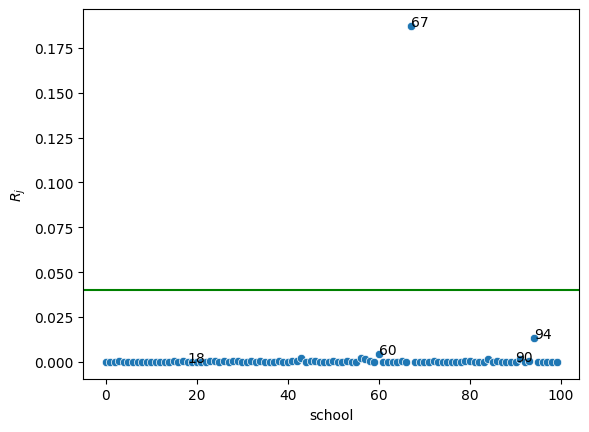}
    \caption{Influential schools based on random-effects Cook's distance with green line as the rule-of-thumb cutoff.}
    \label{fig:r_j} 
     \end{subfigure}
     
\caption{Cook's distance based on random effects}
\label{fig:R}
\end{figure}

Employing the rule-of-thumb cutoff for Cook's distance \citep{Bollinger1981} of $4/n_\text{stu}$ as well as $4/n_\text{sch}$ for determining if a observation is influential due to the fact that the $\chi^2$ reference distributions are too conservative. From Figure~\ref{fig:c_i}, the three student outliers are flagged by the fixed-effects Cook's distance. However, many other students are flagged as well (28 in total), indicating that a more conservative cutoff could be used \citep{loy2014hlmdiag}; however for consistency we choose the same cutoff for all metrics. From Figure~\ref{fig:c_j} three schools of the five are correctly identified for fixed-effects Cook's distance. This is expected due to some outlier schools are not as extreme as others. For the random-effects Cook's distance (Figure~\ref{fig:r_i}) no student outliers are identified due to the fact that the students shouldn't have much of an effect on $\sigma^2_b$ and it may be difficult to have an effect on $\sigma^2$. Similarly, only one school is identified for random-effects Cook's distance (Figure~\ref{fig:r_j}).

\subsection{Linear MEIOM}
The Linear MEIOM for both the student-level and school-level hierarchies is computed with confidence levels for the Jarque-Bera tests set to 0.01, the Stouffer test set to 0.05 and the final Cram\'er-von Mises (CvM) test set to 0.05. The lower confidence levels for the Jarque-Bera test are used so that it is harder to reject per mode normality while overall, the normality and $\chi^2$ tests are easier to reject. These values will be used throughout the paper.

\subsubsection{Student level}
\begin{figure}
 \begin{subfigure}[b]{0.490\textwidth}
         \includegraphics[scale=0.350]{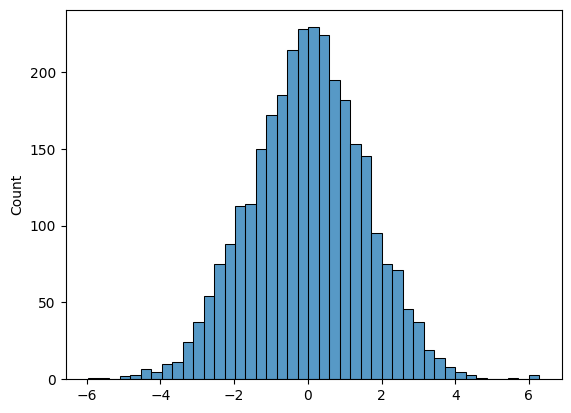}
    \caption{Student-level SHAP values}
        \label{fig:x_shap} 
     \end{subfigure}
 \begin{subfigure}[b]{0.490\textwidth}
    \includegraphics[scale=0.350]{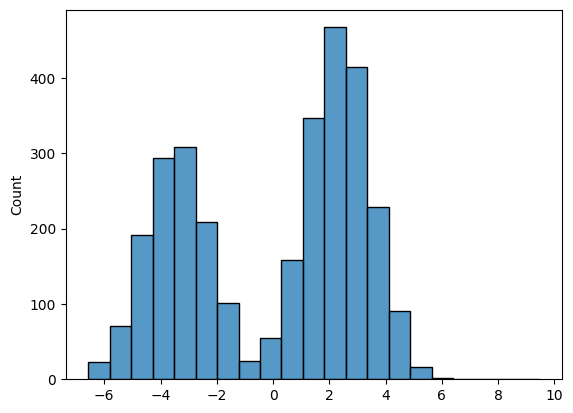}
    \caption{Student-level residuals}
    \label{fig:student_resid} 
     \end{subfigure}
     
\caption{Distributions of linear student-level SHAP values and residuals.}
\label{fig:linear_meiom_student}
\end{figure}

\begin{figure}
    \centering
    \includegraphics[scale=0.350]{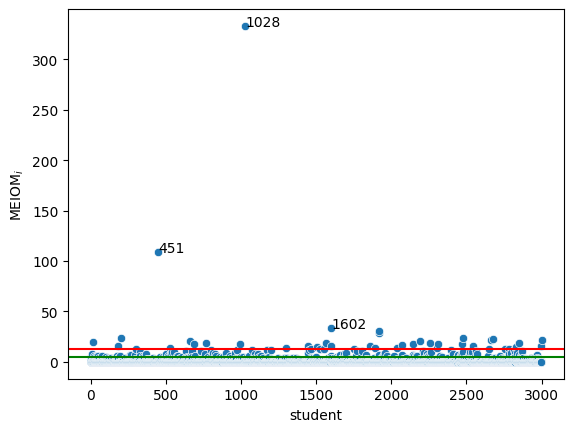}
    \caption{Linear student-level MEIOM where green and red lines represent the $\alpha=0.05$ and $\alpha=0.01$ cut-offs.}
    \label{fig:linear_iom_i}
\end{figure}

Figure \ref{fig:linear_meiom_student} plots the SHAP values and residuals for the student-level covariate and the student-level residuals. The SHAP values appear more unimodal compared with the more bimodal appearance from the residuals. Table \ref{tab:example_linear_student} presents the parameters and results for the SHAP values and residuals. The regularization parameter $\lambda$ is selected so that there is no reason to reject the goodness-of-fit tests. The NFIC selects the unimodal mixture for the SHAP values (0.897) and the bimodal mixture for the residuals (2.23). From Figure \ref{fig:linear_iom_i} the three student outliers are identified; however,  note that a more conservative cutoff should also be used because 41 outliers are also identified at the $\alpha=0.01$ confidence level.

\begin{table}
    \centering
    \begin{tabular}{lcccc}
    \toprule
         &  \textbf{SHAP Unimodal} & \textbf{SHAP Bimodal} & \textbf{Residual Unimodal} & \textbf{Residual Bimodal}\\
        \hline
        $\mu_i$ & 0.0789 & $\{-2.97 2.97\}$ & -0.000800 & $\{-3.29, 3.36\}$ \\
        $\sigma_i$ & 1.05 & $\{1.02, 1.02\}$ & 1.03 & $\{1.14, 1.14\}$\\
        $w_k$ & 1 & $\{0.478, 0.522\}$ & 1 & $\{0.418, 0.582\}$ \\
        Transformations & 6 & 6 & 12 & 4 \\
        Hidden units & 6 & 6 & 12 & 4\\
        Hidden layers & 2 & 2 & 2 & 2\\
        Epochs & 500 & 500 & 1000 & 1000\\
        Tail bound & 7 & 7 & 10 & 10\\
        Number of bins & 8  & 14 & 8 & 16\\
       $\lambda$ &  0.0970 & 0.0498 & 0.0256 & 14.4  \\
       $\hat{p}_{\text{JB}}$ & 0.840 & $\{ 0.0366, 0.233\}$ & 0.875 & $\{0.347, 0.0271\}$\\
       $\hat{p}_{\text{St}}$ & 0.840 & 0.0374 &  0.875 & 0.0506\\
       $\hat{p}_{\text{CvM}}$ & 0.269 & 0.926 & 0.779 & 0.714\\
       NFIC & 0.897 & 4.82 & 4.97 & 2.23 \\
       \bottomrule
        
    \end{tabular}
    \caption{Parameter values, $p$-values and NFIC for the linear unimodal and bimodal SHAP values and residuals.}
    \label{tab:example_linear_student}
\end{table}

\subsubsection{School level}
\begin{figure}
 \begin{subfigure}[b]{0.490\textwidth}
         \includegraphics[scale=0.350]{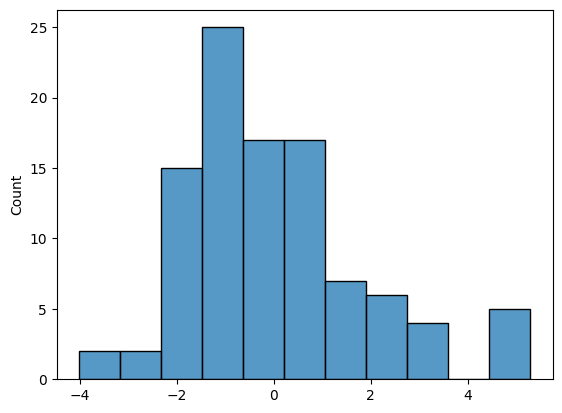}
    \caption{School-level SHAP values}
        \label{fig:school_shap} 
     \end{subfigure}
 \begin{subfigure}[b]{0.490\textwidth}
    \includegraphics[scale=0.350]{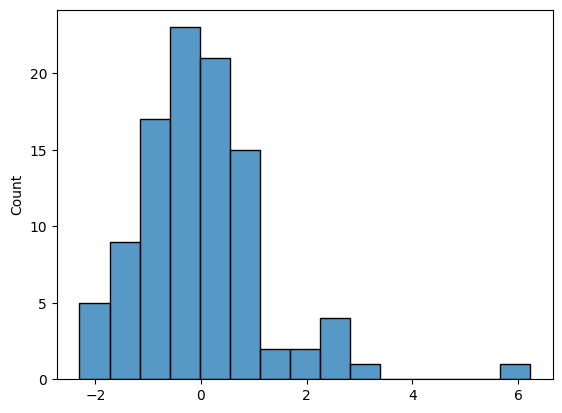}
    \caption{School-level residuals}
    \label{fig:school_resid} 
     \end{subfigure}
     
\caption{Distributions of linear school-level SHAP values and residuals.}
\label{fig:linear_meiom_school}
\end{figure}

Fit the school-level linear MEIOM. Figure \ref{fig:linear_meiom_school} plots the SHAP values and residuals for the school-level covariate and the school-level residuals. Table \ref{tab:example_linear_school} presents the parameters and results for the SHAP values and residuals. The regularization parameter $\lambda$ is selected so that there is no reason to reject the goodness-of-fit tests. Note that the bimodal mixture was not considered due to the small sample size which results in the movement to a single mode. From Figure \ref{fig:linear_iom_j}, three of the 5 outliers are identified at the $\alpha=0.01$ confidence level ($\xi_{1,0.01}=13.0$) with the more egregious ones having a higher MEIOM score. 

\begin{table}
    \centering
    \begin{tabular}{lcc}
    \toprule
         &  \textbf{SHAP Unimodal} & \textbf{Residual Unimodal} \\
        \hline
        $\mu_i$ & 0.0318 &  0.0680  \\
        $\sigma_i$ & 1.08 & 1.10\\
        $w_k$ & 1 &  1 \\
        Transformations & 6 & 6  \\
        Hidden units & 6 & 6 \\
        Hidden layers & 2 & 2 \\
        Epochs & 500 & 500 \\
        Tail bound & 6 & 7 \\
        Number of bins & 12 & 10 \\
       $\lambda$ &  1.40 & 76.2  \\
       $\hat{p}_{\text{JB}}$ & 0.325 &  0.0663 \\
       $\hat{p}_{\text{St}}$ & 0.325 & 0.0663 \\
       $\hat{p}_{\text{CvM}}$ & 0.134 & 0.553\\
       \bottomrule
        
    \end{tabular}
    \caption{Parameter values and $p$-values for the linear unimodal SHAP values and residuals.}
    \label{tab:example_linear_school}
\end{table}

\begin{figure}
    \centering
    \includegraphics[scale=0.350]{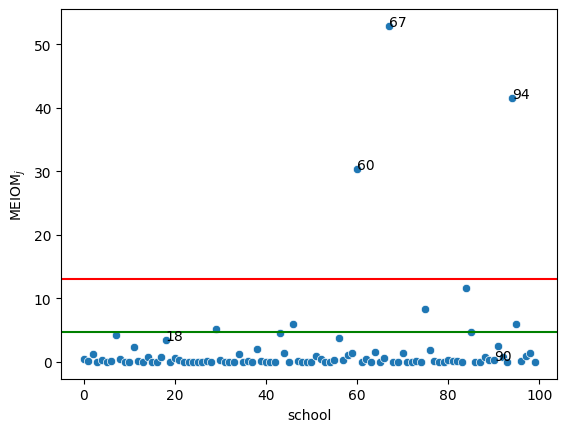}
    \caption{Linear school-level MEIOM where green and red lines represent the $\alpha=0.05$ and $\alpha=0.01$ cut-offs.}
    \label{fig:linear_iom_j}
\end{figure}

\subsection{Mixed Effect Random Forest MEIOM}
In this section, MERF MEIOM is calculated for both the student-level and school-level hierarchies using the same dataset from the generated example.

\subsubsection{Student level}
\begin{figure}
 \begin{subfigure}[b]{0.490\textwidth}
         \includegraphics[scale=0.350]{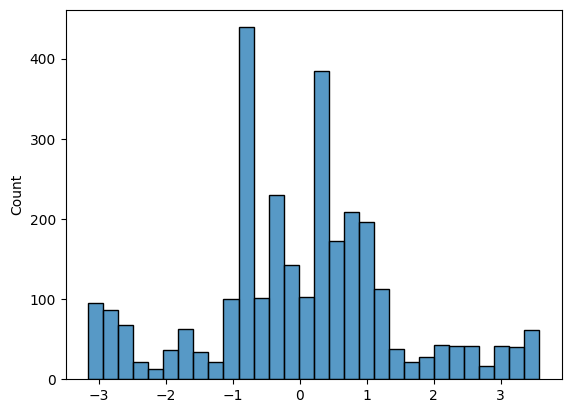}
    \caption{Student-level SHAP values}
        \label{fig:rf_shap} 
     \end{subfigure}
 \begin{subfigure}[b]{0.490\textwidth}
    \includegraphics[scale=0.350]{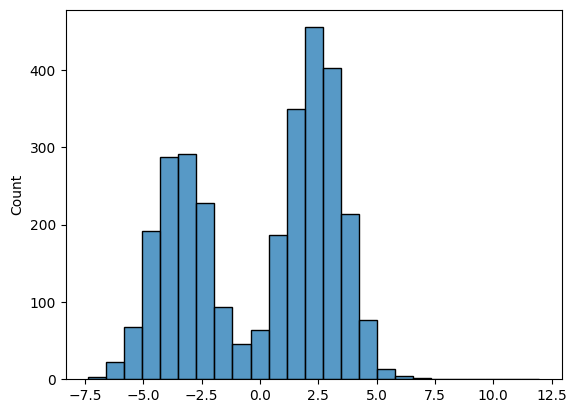}
    \caption{Student-level residuals}
    \label{fig:rf_student_resid} 
     \end{subfigure}
     
\caption{Distributions of MERF student-level SHAP values and residuals.}
\label{fig:rf_meiom_student}
\end{figure}

\begin{figure}
    \centering
    \includegraphics[scale=0.350]{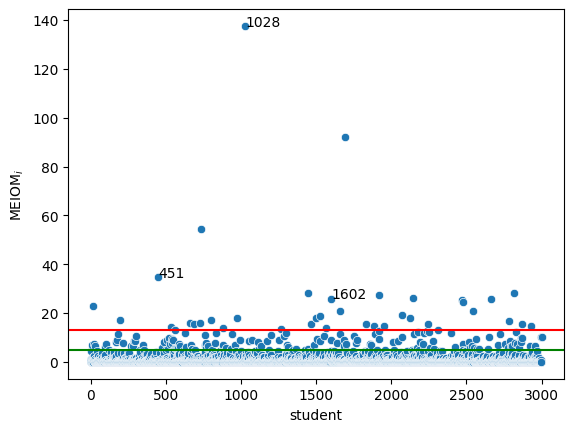}
    \caption{MERF student-level MEIOM where green and red lines represent the $\alpha=0.05$ and $\alpha=0.01$ cut-offs.}
    \label{fig:merf_iom_i}
\end{figure}

Figure \ref{fig:rf_meiom_student} plots the SHAP values and residuals for the student-level covariate and the student-level residuals. Clearly, the SHAP values appear more unimodal compared with the more bimodal appearance from the residuals. Table \ref{tab:example_merf_student} presents the parameters and results for the SHAP values and residuals. The regularization parameter $\lambda$ is selected so that there is no reason to reject the goodness-of-fit tests. The NFIC selects the unimodal mixture for the SHAP values (0.896) and the bimodal mixture for the residuals (3.43). From Figure \ref{fig:merf_iom_i} the three student outliers are identified; however, a more conservative cutoff could also be used because 28 plus 9 additional outliers are also identified at the $\alpha=0.01$ confidence level. There also appears to be two extra outliers with high MEIOM which is different from the linear fit: the higher one corresponding to a outlier that has a large transformed residual and the other one has both a high transformed SHAP value and large transformed residual.

\begin{table}
    \centering
    \begin{tabular}{lcccc}
    \toprule
         & \textbf{SHAP Unimodal} & \textbf{SHAP Bimodal} & \textbf{Residual Unimodal} & \textbf{Residual Bimodal}\\
        \hline
        $\mu_i$ & 0.00740 & $\{-2.95, 2.95\}$ & 0.00980 & $\{-3.97, 3.98\}$ \\
        $\sigma_i$ & 1.01 & $\{0.998, 0.997\}$ & 1.01 & $\{1.08, 1.10\}$\\
        $w_k$ & 1 & $\{0.504, 0.496\}$ & 1 & $\{0.419, 0.581\}$ \\
        Transformations & 6 & 6 & 12 & 6 \\
        Hidden units & 6 & 6 & 12 & 6\\
        Hidden layers & 4 & 4 & 2 & 2\\
        Epochs & 500 & 1000 & 1000 & 1000\\
        Tail bound & 4 & 6 & 13 & 13\\
        Number of bins & 24 & 32 & 32 & 32\\
        $\lambda$ & 0.0970 & 0.0256 & 0.00 & 0.717 \\
        $\hat{p}_{\text{JB}}$ & 0.119 & $\{0.467, 0.354\}$ & 0.143 & $\{0.466, 0.169\}$\\
        $\hat{p}_{\text{St}}$ & 0.119 & 0.373 & 0.143 & 0.230\\
        $\hat{p}_{\text{CvM}}$ & 0.189 & 0.132 & 0.992 & 0.122\\
        NFIC & 0.896 & 5.26 & 5.21 & 3.43 \\
        \bottomrule
    \end{tabular}
    \caption{Parameter values, $p$-values and NFIC for the MERF unimodal and bimodal SHAP values and residuals.}
    \label{tab:example_merf_student}
\end{table}

\subsubsection{School level}
\begin{figure}
 \begin{subfigure}[b]{0.490\textwidth}
         \includegraphics[scale=0.350]{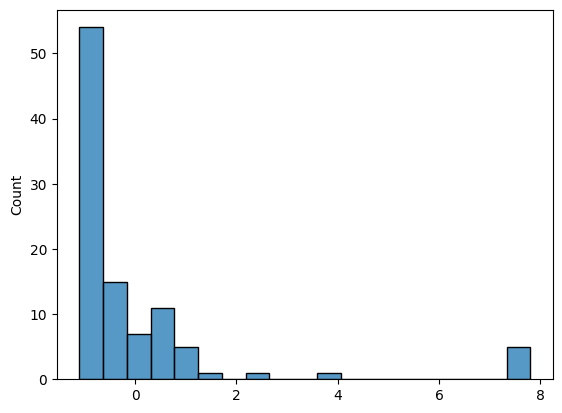}
    \caption{School-level SHAP values}
        \label{fig:rf_school_shap} 
     \end{subfigure}
 \begin{subfigure}[b]{0.490\textwidth}
    \includegraphics[scale=0.350]{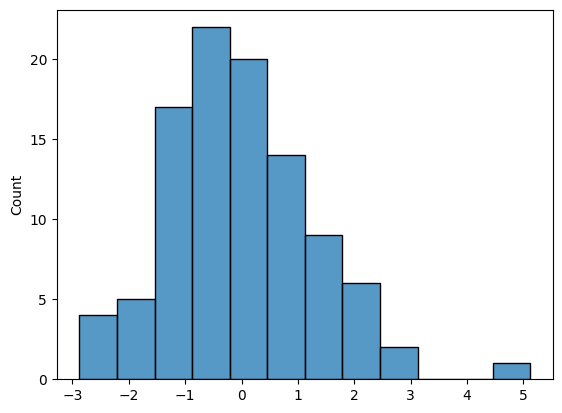}
    \caption{School-level residuals}
    \label{fig:rf_school_resid} 
     \end{subfigure}
     
\caption{Distributions of MERF school-level SHAP values and residuals.}
\label{fig:rf_meiom_school}
\end{figure}

Figure \ref{fig:rf_meiom_school} plots the SHAP values and residuals for the school-level covariate and the school-level residuals. Table \ref{tab:example_rf_school} presents the parameters and results for the SHAP values and residuals. The regularization parameter $\lambda$ is selected so that there is no reason to reject the goodness-of-fit tests. Note that the bimodal mixture was not considered due to the low sample size. From Figure \ref{fig:linear_iom_j}, four of the 5 outliers are identified at $\alpha=0.01$ confidence level ($\xi_{1,0.01}=13.0$) the with the more egregious ones having a higher MEIOM score. Note that outlier 18 is identified in this model and is not in the linear fit. Also, one extra outlier gets marginally identified at the 0.01 level due to both a moderately high transformed SHAP value and transformed residual.

\begin{table}
    \centering
    \begin{tabular}{lcc}
    \toprule
         &  \textbf{SHAP Unimodal} & \textbf{Residual Unimodal}\\
        \hline
        $\mu_i$ & 0.0534 & 0.133 \\
        $\sigma_i$ & 1.04 & 1.13\\
        $w_k$ & 1 & 1 \\
        Transformations & 4 & 4 \\
        Hidden units & 4 & 4 \\
        Hidden layers & 2 & 2 \\
        Epochs & 500 & 500 \\
        Tail bound & 9 & 6 \\
        Number of bins & 32 & 8 \\
        $\lambda$ & 0.0970 & 403 \\
        $\hat{p}_{\text{JB}}$ & 0.141 & 0.326 \\
        $\hat{p}_{\text{St}}$ & 0.141 & 0.326 \\
        $\hat{p}_{\text{CvM}}$ & 0.987 & 0.642\\
        \bottomrule
    \end{tabular}
    \caption{Parameter values and $p$-values for the MERF unimodal school SHAP values and residuals.}
    \label{tab:example_rf_school}
\end{table}

\begin{figure}
    \centering
    \includegraphics[scale=0.350]{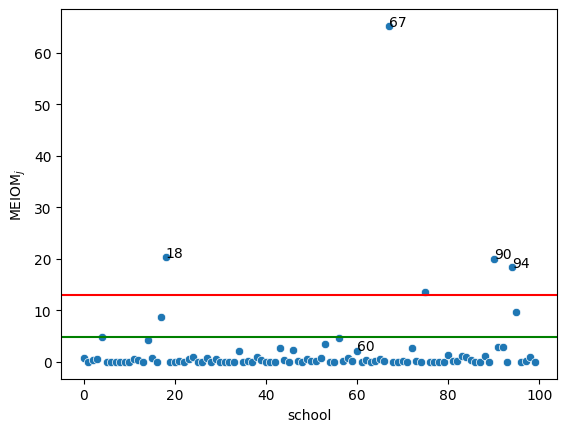}
    \caption{MERF school-level MEIOM where green and red lines represent the $\alpha=0.05$ and $\alpha=0.01$ cut-offs.}
    \label{fig:linear_iom_j}
\end{figure}

\section{Application}\label{application}
The Programme for International Student Assessment\footnote{https://www.oecd.org/en/data/datasets/pisa-2022-database.html} is a large-scale international assessment conducted by the OECD to evaluate the skills of approximately 15-year-old students across participating countries and economies. PISA focuses primarily on mathematics, reading, and science literacy, emphasizing students’ ability to apply knowledge to real-world problems rather than simply recall curriculum content. The 2022 data set is used. 

The PISA data have a hierarchical structure, with students nested within schools, making them particularly suitable for multilevel and mixed-effects analyses. In addition to achievement measures, PISA includes extensive student- and school-level background information, such as socioeconomic characteristics, attitudes toward learning, school resources, and educational practices. For this paper, consider data from Singapore (6559 students and 164 schools)  and the Philippines (7167 students 188 schools). We use several abbreviations to name variables and covariates (Table \ref{tab:abbreviations}).

\begin{table}[]
    \centering
    \caption{Abbreviations for PISA data}
    \label{tab:abbreviations}
    \begin{tabular}{ll}
     \toprule
\textbf{Abbreviation} & \textbf{Full text} \\
\midrule
PHL & Philippines \\
SGP & Singapore\\
ESCS & Index of economic, social and cultural status\\
PV1MATH & Mathematics score\\
MATHPERS & Effort and persistence in mathematics\\
MATHEFF & Mathematics self-efficacy: formal and applied mathematics \\
PROPMATH & Proportion of mathematics teachers at school\\

\bottomrule
    \end{tabular}
\end{table}

\subsection{Small model}
The small model does not explicitly account for the country effects or school-level covariates; we assume that school-level effects will adequately describe the two countries. This omission implies that country level variability is already accounted for by the school level effects. If we are incorrect, then the bimodal MEIOM may be indicated for use. There are a total of 13726 students in 352 schools. We regress,

\begin{equation}
    \mathrm{PV1MATH}_{ij} = f(\mathrm{ESCS}_{ij}; \bm{\theta}_{\text{small}}) + b_{j} + \varepsilon_{ij},
\end{equation}
where $b_{j}$ is the random intercept for school $j$. A MEGB model \citep{olaniran2025mixed} is fitted using an 80/20 stratified 3-fold testing split to tune hyperparameters (32 iterations).  The root mean squared error (RMSE) is 71.3 on the test set. 

\subsubsection{Student level}
\begin{figure}
 \begin{subfigure}[b]{0.490\textwidth}
 \centering
        \includegraphics[scale=0.350]{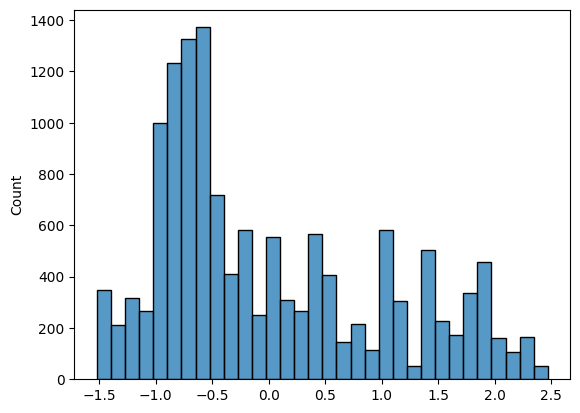}
        \caption{SHAP values (standardized).}
        \label{fig:megb_shap_small} 
     \end{subfigure}
 \begin{subfigure}[b]{0.490\textwidth}
 \centering
    \includegraphics[scale=0.350]{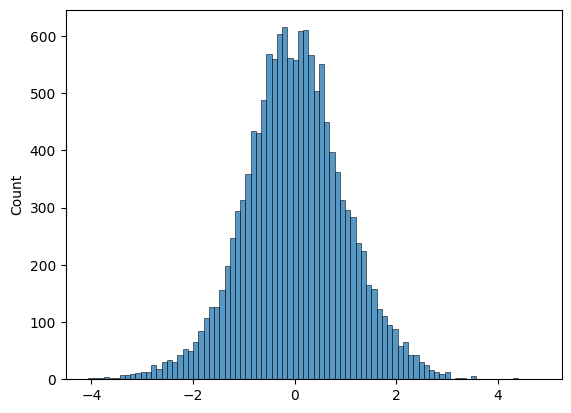}
    \caption{Residuals (standardized).}
    \label{fig:megb_resid_small} 
     \end{subfigure}
     
\caption{Distributions of MEGB student-level SHAP values and residuals for the small model.}
\label{fig:megb_stu_small}
\end{figure}

At the student level, there is no clear separation of the countries (Figure \ref{fig:megb_resid_small}). For the student-level SHAP values (Figure \ref{fig:megb_shap_small}), we use 13 transformations, 13 hidden units, 2 layers, a tail bound of 3, 10 bins, 1000 epochs, a unimodal base distribution with mean $0.0740$, standard deviation 0.851 and tune $\lambda=3.06\times10^{-7}$. For the student-level residuals we use 4 transformations, 4 hidden units, 2 layers, a tail bound of 3, 8 bins, 500 epochs, a unimodal base distribution with mean $-0.0952$, standard deviation 1.05 and tune $\lambda=0.0131$. We do not reject any of the hypothesis tests for the NF. 137 outliers are reported at the $\alpha=0.01$ ($\xi_{1,0.01}=13.0$) level along with the top 5 outliers (Table \ref{tab:megb_iom_stu_small}). The transformed residuals play an important role in the identification of the first outlier while the other top four are more of a combination of the transformed SHAP value and the transformed residual. 

\begin{table}[ht]
\centering
\caption{Top 5 outlier students determined by the MEGB IOM.}
\label{tab:megb_iom_stu_small}
\begin{tabular}{cccccc}
\toprule
\textbf{Country} & \textbf{PV1MATH} & \textbf{ESCS} & \textbf{IOM} & \textbf{SHAP (as a $\chi^2_1$)} & \textbf{Residuals (as a $\chi^2_1$)} \\\\
\midrule
SGP & 364 & 1.09  & 87.4 & 6.16 & 14.2 \\\\
SGP & 454 & 1.36  & 62.9 & 7.98 & 7.88 \\\\
SGP & 386 & 1.35  & 45.1 & 7.98 & 5.65 \\\\
PHL & 524 & -0.890 & 44.9 & 5.53 & 8.13 \\\\
PHL & 498 & -0.730 & 44.3 & 6.46 & 6.86 \\\\
\bottomrule
\end{tabular}
\end{table}

\subsubsection{School level}
\begin{figure}
    \centering
    \includegraphics[scale=0.350]{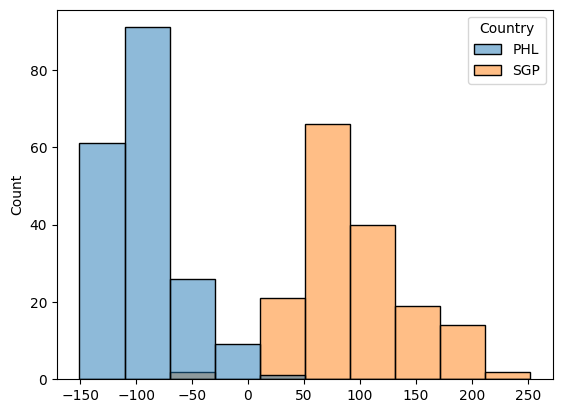}
    \caption{Estimated random intercepts for Singapore and the Philippines.}
    \label{fig:bhat_small}
\end{figure}

First, a unimodal base distribution is fit with mean 0.102, standard deviation 87.7, 2 transformations, 2 hidden units, 2 layers, a tail bound of 250, 16 bins, 500 epochs, and tune $\lambda=0.0970$. Next a bimodal base distribution is fit with means $\{-99.8, 100.0\} $, standard deviations $\{44.8, 51.7\}$, weights $\{0.502, 0.498\}$, 2 transformations, 2 hidden units, 2 layers, a tail bound of 250, 16 bins, 500 epochs, and tune $\lambda=0.189$. The same set of previous hypothesis tests are not rejected. The unimodal NFIC is 880 whereas the bimodal NFIC is 155, clearly selecting the bimodal specification and provides evidence for a model that has latent variable structure. There is a clear separation of the countries (Figure \ref{fig:bhat_small}). Due to the absence of school-level covariates, only the random intercept residual is used and when transformed, becomes a $\chi^2_1$ random variable. Therefore, the cutoff is $\chi^2_1(0.99)=6.63$. The two labeled outliers for the school-level are from Singapore and have very large transformed random intercepts.

\begin{table}[ht]
\centering
\caption{Outlier school exceeding the $\chi^2_1(0.99)$ quantile.}
\label{tab:school_estimates}
\begin{tabular}{ccc}
\toprule
\textbf{Country} & \textbf{Residuals} & \textbf{Residuals (as a $\chi^2_1$)} \\
\midrule
SGP & 252  & 8.81 \\
SGP & 240  & 7.48 \\

\bottomrule
\end{tabular}
\end{table}

\subsection{Full model}
The full model explicitly accounts for the country effects as both a fixed effect and as context for the NF. There are a total of 12955 students in 352 schools. The chosen covariates are described in the relation,

\begin{equation}
    \mathrm{PV1MATH}_{ijk} = f(\mathrm{ESCS}_{ijk}, \mathrm{MATHPERS}_{ijk}, \mathrm{MATHEFF}_{ijk}, \mathrm{PROPMATH}_{jk}, C_k; \bm{\theta}_{\text{full}}) + b_{jk} + \varepsilon_{ijk},
\end{equation}
where $C_k$ denotes country as a categorical fixed-effect covariate. A MEGB model is fitted and using an 80/20 stratified 3-fold testing split to tune hyperparameters (32 iterations), with 64.1 RMSE on the test set. 

\subsubsection{Student level}
At the student level, the SHAP values show strong  multimodality in Figure \ref{fig:megb_shap_full} but bell-shaped residuals in Figure \ref{fig:megb_resid_full}. Due to the many modes and the higher dimensionality, unimodal IOM is used on the SHAP values. In particular, we use 12 transformations, 12 hidden units, 2 layers, a tail bound of 4, 13 bins, 1000 epochs, a unimodal base distribution with mean vector $[-0.0342,  0.00270, -0.0577]'$, diagonal covariance matrix $\mathrm{diag}(0.906^2, 0.864^2, 0.892^2)$ and tune $\lambda=0.00674$. For the student-level residuals we use 12 transformations, 12 hidden units, 3 layers, a tail bound of 5.10, 8 bins, 500 epochs, a unimodal base distribution with mean $0.00670$, standard deviation 1.0335 and tune $\lambda=3.059\times 10^{-7}$. None of the hypothesis tests for the NF are rejected. A total of 132 outliers are identified at the $\alpha=0.01$ ($\xi_{3,0.01}=28.7$) with the top 5 outliers shown in Table \ref{tab:megb_iom_stu_small}. The transformed residual portion dominates the first and fifth labeled outlier whereas the transformed SHAP values play a bigger role in the second and fourth labeled outlier. The other top five outliers have a combination of transformed SHAP and residual outlying behaviour. 

\begin{figure}
 \begin{subfigure}[b]{0.490\textwidth}
 \centering
        \includegraphics[scale=0.400]{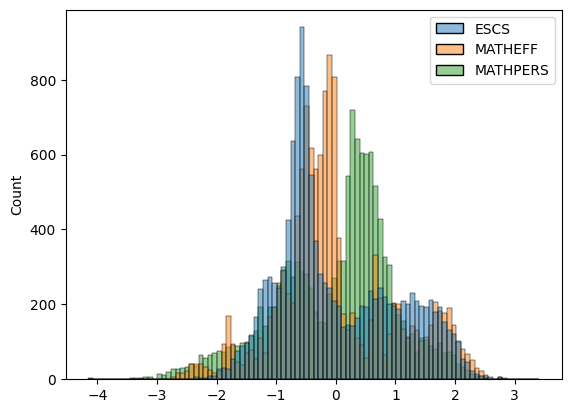}
        \caption{SHAP values (standardized).}
        \label{fig:megb_shap_full} 
     \end{subfigure}
 \begin{subfigure}[b]{0.490\textwidth}
 \centering
    \includegraphics[scale=0.400]{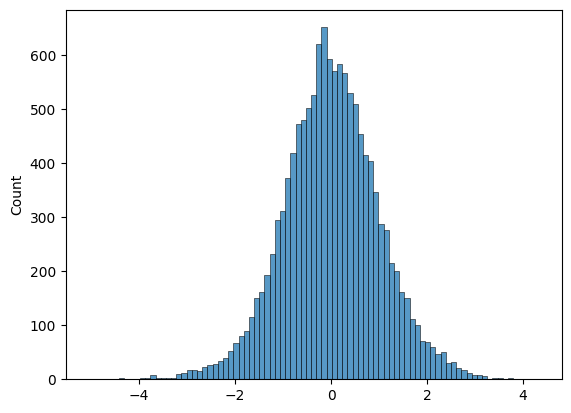}
    \caption{Residuals (standardized).}
    \label{fig:megb_resid_full} 
     \end{subfigure}
     
\caption{Distributions of MEGB student-level (marginal and standardized) SHAP values and residuals for the full model.}
\label{fig:megb_stu_small}
\end{figure}

\begin{table}[ht]
\centering
\caption{Top 5 outlier students determined by the MEGB IOM full model.}
\label{tab:megb_iom_stu_full}
\begin{tabular}{cccccc}
\toprule
\textbf{Country} & \textbf{PV1MATH} & \textbf{ESCS} & \textbf{MEIOM} & \textbf{SHAP (as a $\chi^2_3$)} & \textbf{Residuals (as a $\chi^2_1$)} \\
\midrule
SGP & 943 & 1.11   & 127  & 9.17 & 13.8 \\
SGP & 722 & 1.25   & 109  & 22.9 & 4.74 \\
PHL & 263 & -0.114 & 98.2 & 12.1 & 8.12 \\
SGP & 747 & 0.903  & 91.5 & 14.5 & 6.29 \\
PHL & 201 & -3.18  & 87.0 & 8.54 & 10.2 \\
\bottomrule
\end{tabular}
\end{table}

\subsubsection{School level}
At the school level for the full model Figure \ref{fig:megb_resid_full} does not showcase bimodality. However, some outliers reside in the tail of the distribution. The unimodal NF is used for both the school-level SHAP values and residuals. For the SHAP values (Figure \ref{fig:megb_shap_full}), we use 2 transformations, 2 hidden units, 2 layers, a tail bound of 3, 8 bins, 500 epochs, a unimodal base distribution with mean $0.000100$, standard deviation 1.018 and tune $\lambda=10^{10}$\footnote{A large $\lambda$ is used because the data is already very normal.}. For the student-level residuals we use 4 transformations, 4 hidden units, 2 layers, a tail bound of 5.2, 14 bins, 500 epochs, a unimodal base distribution with mean $0.133$, standard deviation 1.1347 and tune $\lambda=0.0131$. No hypothesis tests are rejected for the NF. There are 5 outliers at the $\alpha=0.01$ confidence level ($\xi_{1,0.01}=13.0$) (Table \ref{tab:megb_iom_sch_full}). At the school level, the transformed SHAP values play a large role in the outlying behaviour for the 1st and 3rd outlier schools. The second outlier school is labeled due to a larger transformed residual. The other outliers are a combination of both the transformed SHAP and residual.

\begin{figure}
 \begin{subfigure}[b]{0.490\textwidth}
 \centering
        \includegraphics[scale=0.400]{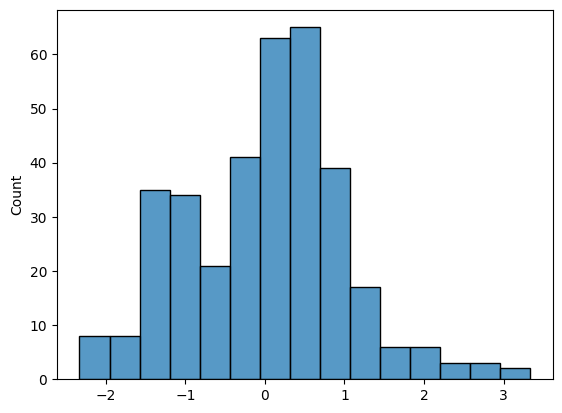}
        \caption{SHAP values (standardized).}
        \label{fig:megb_shap_full} 
     \end{subfigure}
 \begin{subfigure}[b]{0.490\textwidth}
 \centering
    \includegraphics[scale=0.400]{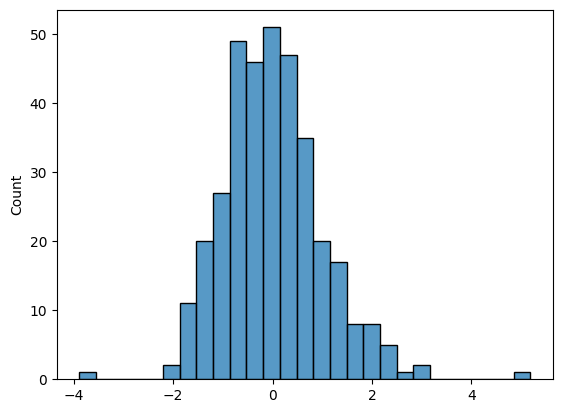}
    \caption{Residuals (standardized).}
    \label{fig:megb_resid_full} 
     \end{subfigure}
     
\caption{Distributions of MEGB school-level (marginal and standardized) SHAP values and residuals for the full model.}
\label{fig:megb_stu_small}
\end{figure}

\begin{table}[ht]
\centering
\caption{Outlier schools determined by the MEGB IOM full model.}
\label{tab:megb_iom_sch_full}
\begin{tabular}{cccccc}
\toprule
\textbf{Country} & \textbf{PROPMATH} & \textbf{PV1MATH} & \textbf{IOM} & \textbf{SHAP (as a $\chi^2_1$)} & \textbf{Residuals (as a $\chi^2_1$)} \\
\midrule
SGP & 0.139 & 728 & 52.1 & 10.5 & 4.98 \\
SGP & 0.196 & 391 & 36.3 & 4.98 & 7.29 \\
SGP & 0.138 & 699 & 26.6 & 10.7 & 2.49 \\
SGP & 0.213 & 743 & 23.0 & 3.49 & 6.60 \\
SGP & 0.197 & 419 & 18.7 & 4.51 & 4.14 \\
\bottomrule
\end{tabular}
\end{table}

\section{Conclusion}\label{conclusion}
This paper extends the IOM to hierarchical data and further expands its applicability through more generalized NF transformations. The MEIOM considers clustered data by subdividing the framework into individual and group-level observations. The result combines hierarchical residual information with level-specific SHAP values, allowing influence to be assessed at multiple levels of the data hierarchy while remaining applicable to both traditional MEMs and mixed-effects machine-learning models.  This contribution facilitates identification influential individuals and also influential groups, clusters, and higher-level units that may unduly impact prediction accuracy or inference. 

The restriction that the NF must map the SHAP values and residuals to a single Gaussian base distribution is relaxed. Allowing a Gaussian-mixture base distribution accommodates multimodal structures that would otherwise require a more complicated transformation to map to a unimodal reference distribution. We further introduce the Normalizing Flow Information Criterion (NFIC) to balance the complexity of the base distribution against the amount of transport required by the NF while providing a diagnostic for potential model fit challenges. In addition, conditional NFs allow available contextual information to influence the transformation while retaining a common calibrated reference distribution. 

The PISA application demonstrates how the MEIOM framework can reveal model misspecification and lead to improved model understanding. When country was omitted from the smaller model, the school-level random effects displayed clear separation between Singapore and the Philippines, and the NFIC strongly favoured a bimodal rather than unimodal representation. Country was then added to the model as a fixed effect and as context within the NF, resulting in a more accurate model in terms of RMSE while also illustrating how the distributions of both student- and school-level SHAP values and residuals changed.

 Although the MEIOM is developed here for random intercept MEM, the MEIOM framework should generalize beyond the two-level random-intercept setting considered here to accommodate random slopes, crossed or nested random effects, and deeper hierarchical structures. Full exploration thereof is left for future work. Additional directions for future work build on the NF extensions. First, the Gaussian-mixture formulation currently relies on sufficiently well-separated components so that observations can be assigned to a single mode and the resulting within-component Mahalanobis distances can be approximated by chi-square distributions. Future work could relax this assumption and develop inference for settings in which mixture components overlap substantially or mode assignment is uncertain. More general base distributions, including Gaussian mixtures with full covariance matrices and potentially non-Gaussian mixture components, could also be investigated.  Finally, the NFIC could be extended to account for additional sources of complexity, including alternative definitions of transport cost.

\bibliography{bib}

\end{document}